# Full- & Reduced-Order State-Space Modeling of Wind Turbine Systems with Permanent-Magnet Synchronous Generator

Christoph M. Hackl[‡,⋆], Martin Pfeifer[†,⋆], Korbinian Schechner[‡], Pol Jané-Soneira[†] and Sören Hohmann[†]

**Abstract**

Wind energy is an integral part of nowadays energy supply and one of the fastest growing sources of electricity in the world today. Accurate models for wind energy conversion systems (WECSs) are of key interest for the analysis and control design of present and future energy systems. Existing control-oriented WECSs models are subject to unstructured simplifications, which have not been discussed in literature so far. Thus, this technical note presents are thorough derivation of a physical state-space model for permanent magnet synchronous generator WECSs. The physical model considers all dynamic effects that significantly influence the system's power output, including the switching of the power electronics. Alternatively, the model is formulated in the $(a, b, c)$- and $(d, q)$-reference frame. Secondly, a complete control and operation management system for the wind regimes II and III and the transition between the regimes is presented. The control takes practical effects such as input saturation and integral windup into account. Thirdly, by a structured model reduction procedure, two state-space models of WECS with reduced complexity are derived: a non-switching model and a non-switching reduced-order model. The validity of the models is illustrated and compared through a numerical simulation study.

CONTENTS

[‡]C.M. Hackl and K. Schechner are with the research group "Control of renewable energy systems" (CRES) at the Munich School of Engineering (MSE), Technische Universität München (TUM), Germany.

[†]M. Pfeifer, P. Jané Soneira and S. Hohmann are with the Institute of Control Systems (IRS) at the Karlsruhe Institute of Technology (KIT), Germany.

[*]The first four authors are in alphabetical order and contributed equally to the paper. Corresponding authors are C.M. Hackl (christoph.hackl@tum.de) and M. Pfeifer (martin.pfeifer@kit.edu).

## NOTATION

$\mathbb{N}, \mathbb{R}, \mathbb{C}$: natural, real, complex numbers. $\boldsymbol{x} := (x_1, \ldots, x_n)^\top \in \mathbb{R}^n$: column vector, $n \in \mathbb{N}$ where "$^\top$" and ":=" mean "transposed" (interchanging rows and columns of a matrix or vector) and "is defined as", resp., $\mathbf{0}_n \in \mathbb{R}^n$: zero vector. $\mathbf{1}_n \in \mathbb{R}^n$: unity vector. $\boldsymbol{A} \in \mathbb{R}^{n \times n}$: (square) matrix with $n$ rows and columns. $\boldsymbol{A}^{-1}$: inverse of $\boldsymbol{A}$ (if exists). $\boldsymbol{A}^{-\top}$: transposed inverse of $\boldsymbol{A}$ (if exists). Scalar and vector saturation, for $b > a$ an $\widehat{u} > 0$, given by

$$\begin{aligned} \mathrm{sat}_a^b \colon \mathbb{R} &\to [a,\, b], \\ x &\mapsto \mathrm{sat}_a^b(x) := \begin{cases} b & , x \geq b \\ x & , a < x < b \\ a & , x \leq a. \end{cases} \end{aligned} \quad \text{and} \quad \begin{aligned} \mathbf{sat}_{\widehat{u}} \colon \mathbb{R}^n &\to \{\boldsymbol{x} \in \mathbb{R}^n \mid \|\boldsymbol{x}\| \leq \widehat{u}\}, \\ \boldsymbol{x} &\mapsto \mathbf{sat}_{\widehat{u}}(\boldsymbol{x}) := \begin{cases} \widehat{u} \dfrac{\boldsymbol{x}}{\|\boldsymbol{x}\|} & , \|\boldsymbol{x}\| \geq \widehat{u} \\ \boldsymbol{x} & , \|\boldsymbol{x}\| < \widehat{u}. \end{cases} \end{aligned} \tag{1}$$

Transition function for $\hat{a} \in \mathbb{R}$ and $\Delta > 0$, given by

$$f_{\hat{a},\Delta} \colon \mathbb{R} \to [0,\, 1], \quad x \mapsto f_{\hat{a},\Delta}(x) := \begin{cases} 0 & , x \geq \hat{a} \\ \frac{-x+\hat{a}}{\Delta} & , \hat{a} - \Delta \leq x < \hat{a} \\ 1 & , x < \hat{a} - \Delta \end{cases} \tag{2}$$

$\mathcal{C}(I; Y)$ space of continuous functions mapping $I \to Y$. Two different reference frames (coordinate systems) will be considered (i) three-phase $(a, b, c)$-reference frame $\boldsymbol{x}^{abc} := (x^a, x^b, x^c)^\top \in \mathbb{R}^3$ and (ii) synchronously rotating $(d, q)$-reference frame with $\boldsymbol{x}^{dq} = (x^d, x^q)^\top \in \mathbb{R}^2$ where $\boldsymbol{x}^{abc}$ and $\boldsymbol{x}^{dq}$ are related by the (reduced) Clarke-Park-transformation for $\kappa \in \left\{\frac{2}{3}, \sqrt{\frac{2}{3}}\right\}$ as follows [1, App. A.5]

$$\boldsymbol{x}^{dq} = \underbrace{\kappa \begin{bmatrix} \cos(\phi_\mathrm{p}) & \cos(\phi_\mathrm{p} - \frac{2\pi}{3}) & \cos(\phi_\mathrm{p} - \frac{4\pi}{3}) \\ -\sin(\phi_\mathrm{p}) & -\sin(\phi_\mathrm{p} - \frac{2\pi}{3}) & -\sin(\phi_\mathrm{p} - \frac{4\pi}{3}) \end{bmatrix}}_{=:\boldsymbol{T}_\mathrm{cp}(\phi_\mathrm{p}) \in \mathbb{R}^{2\times 3}} \boldsymbol{x}^{abc} \quad \text{with} \quad \boldsymbol{T}_\mathrm{cp}(\phi_\mathrm{p})^{-1} := \frac{2}{3\kappa} \begin{bmatrix} \cos(\phi_\mathrm{p}) & -\sin(\phi_\mathrm{p}) \\ \cos(\phi_\mathrm{p} - \frac{2\pi}{3}) & -\sin(\phi_\mathrm{p} - \frac{2\pi}{3}) \\ \cos(\phi_\mathrm{p} - \frac{4\pi}{3}) & -\sin(\phi_\mathrm{p} - \frac{4\pi}{3}) \end{bmatrix} \in \mathbb{R}^{3\times 2}. \tag{3}$$

Rotation matrices (counter-clock wise rotation by $\frac{\pi}{2}$), given by

$$\boldsymbol{J} := \begin{bmatrix} 0 & -1 \\ 1 & 0 \end{bmatrix} \quad \text{and} \quad \boldsymbol{J}_\Sigma := \frac{1}{\sqrt{3}} \begin{bmatrix} 0 & -1 & 1 \\ 1 & 0 & -1 \\ -1 & 1 & 0 \end{bmatrix} \tag{4}$$

## I. INTRODUCTION

Sustainable electrical energy is a major concern of modern society. Wind power represents a renewable and carbon-free energy resource which can be made available on a large scale by wind energy conversion systems (WECSs). During the last two decades, electricity generation by wind power experienced a vast expansion leading to a global cumulative installed generation capacity of about 486.8 GW in 2016 [2]. A WECS is a complex system, which covers multiple physical domains, as aerodynamical, mechanical and electrical subsystems. Due to this complexity, studies which have to incorporate the behavior of WECSs mostly apply model-based methods, where the starting point is the derivation of suitable WECS model. There have been many simulation studies investigating the impact of WECSs to power systems [3]–[10]. Moreover, WECS models have been used for real-time hardware-in-the-loop simulations [11], [12]. Most notable, the existence of models is indispensable for the design of control and state estimation methods for WECSs [13]–[21]. As the validity of the models plays a crucial role for these studies, modeling of WECS is a considerable aspect of nowadays power system research.

The history of WECS modeling goes back to the end of the 1970s, where first wind power impact studies were undertaken [22]. In the 1980s, first WECS simulation models for fixed-speed WECSs were presented [23] and applied to large scale transient stability computer programs [3]. Since this time, numerous models of fixed-speed and variable-speed WECSs are noted. These models can be classified according to various criteria: (a) generic models [3]–[21], [24]–[42] and manufacturer-specific models [23], [43]; (b) induction generator WECS models [4], [9], [21], [25], [27], [38], [39], doubly fed induction generator WECS models [6], [7], [10], [12], [16], [20], [26], [29]–[31], [35], [43] and synchronous generator WECS models [3], [8], [9], [11], [12], [15], [19], [24]–[26], [28], [33], [34]; (c) high-order models with deep physical insight [8], [12], [15], [16], [19]–[21], [24]–[26] and reduced-order models, where at least one subsystem of the WECS is significantly simplified [13]–[15], [27], [28]; (d) simulation models [3]–[10], [24], [25], [28]–[34], [39], [40], [43] and control design models [13]–[18], [21], [25], [27], [35]–[37], [41]. The latter are characterized by a closed form mathematical representation in state space.

The publication at hand aims at the development of generic control design models for variable-speed variable-pitch synchronous generator WECSs. In literature, there exists a considerable number of publications addressing the derivation of high-order models for this system type. In [12], [21] nonlinear state space models of WECSs are presented. However, the switching behavior of the power electronics components are not considered explicitly. In [11], [24] simulation models which precisely describe the dynamics of the WECS, including the switching of the power electronics in the synchronous reference frame, are presented. However, the models are not given in closed-form representation which hampers their application to control design. In [8], [25], [44] dynamic simulation models of variable-speed synchronous generator WECSs are presented. However, the switching of the power electronics is not considered explicitly and the models are not given in closed-form representation.

Besides high-order models, there exist numerous reduced-order models for variable-speed synchronous generator WECSs in literature. A common approach to model reduction is to simplify the dynamics of certain subsystems, as for e.g. the generator [9], [13], [17], [43] or the power electronics [13], [17], [27], [33]. In [14], [15], [18], [27], [36] reduced-order linear state space models of WECSs are presented. It must be noted, that there exists no study which discusses the effects of simplifying certain subsystems to the validity of the overall WECS models. In consequence, the validity and usability of the reduced-order models remains vague.

In conclusion, to the best of the authors' knowledge, there exists no high-order control design model for variable-speed WECSs which includes all relevant dynamic effects, in particular with respect to the switching of the power electronics and the pitch control system. Moreover, there exists no structured WECS model reduction approach with a discussion of the validity of the reduction steps. This publication tries to fill this gap by introducing control design models for the dynamic relation between the input wind speed and produced electrical output in variable-speed synchronous generator WECSs. The three main contributions are: (a) a first formulation and a detailed derivation of a thorough state-space model of a variable-speed WECS, which captures all dynamic processes that significantly affect the power output of the system (including the switching of the power electronics and modulation schemes), (b) a complete description of the underlying (cascaded) control and operation management systems with a consideration of practical constraints as saturation and integral windup, and (c) a structured derivation of reduced-complexity state-space models of a variable-speed WECSs with an investigation on the validity of the reduction steps. The modeling is subdivided into the modeling of the physical system and the modeling of the control systems and operation management. The models focus on the operating regions II and III with wind speeds varying between cut-in and cut-out wind speed.

## II. System Description

The considered wind energy conversion system is shown in Fig. 1 and represents a state of the art WECS: a variable-speed, variable-pitch, three bladed, horizontal axis, lift turbine in up-wind position. Only a single wind turbine is considered, without aerodynamical interaction between multiple turbines as described by the wake effect. Regarding the type of generator (in our case permanent magnet synchronous machine (PMSM)), the WECS might or might not comprise a gear between turbine and generator. The generator feeds the converted power through a full-scale back-to-back converter and grid-side filter to the point of common coupling (PCC). The transformer is not explicitly modelled, but could easily be added. The grid is assumed to be symmetrical and stiff, so that the grid-side voltage source inverter (VSI) operates in grid-feeding mode. So an electrical interaction between multiple WECS is not considered.

Depending on the actual wind speed $v_{\mathrm{w}}$ (in $\frac{\mathrm{m}}{\mathrm{s}}$), the wind turbine system will operate in one of the four regimes of operation (see Fig. 2). For too less or too much wind (i.e. $v_{\mathrm{w}} < v_{\mathrm{w,cut-in}}$ in regime I and $v_{\mathrm{w,cut-out}} \leq v_{\mathrm{w}}$ in regime IV, resp.), the wind turbine is (usually[1]) at standstill or in idle mode: The turbine angular velocity is zero, i.e. $\omega_{\mathrm{t}} = 0\,\frac{\mathrm{rad}}{\mathrm{s}}$ or the machine torque is zero, i.e. $m_{\mathrm{m}} = 0\,\mathrm{N\,m}$, hence the turbine (output) power is $p_{\mathrm{t}} = 0\,\mathrm{W}$. In regime II, the wind speed is below the nominal wind

[1] Some companies use more sophisticated control methods for high winds, e.g. see patent [45] of REpower Systems for a reduced power production above $v_{\mathrm{w,cut-out}}$ instead of a shut down.

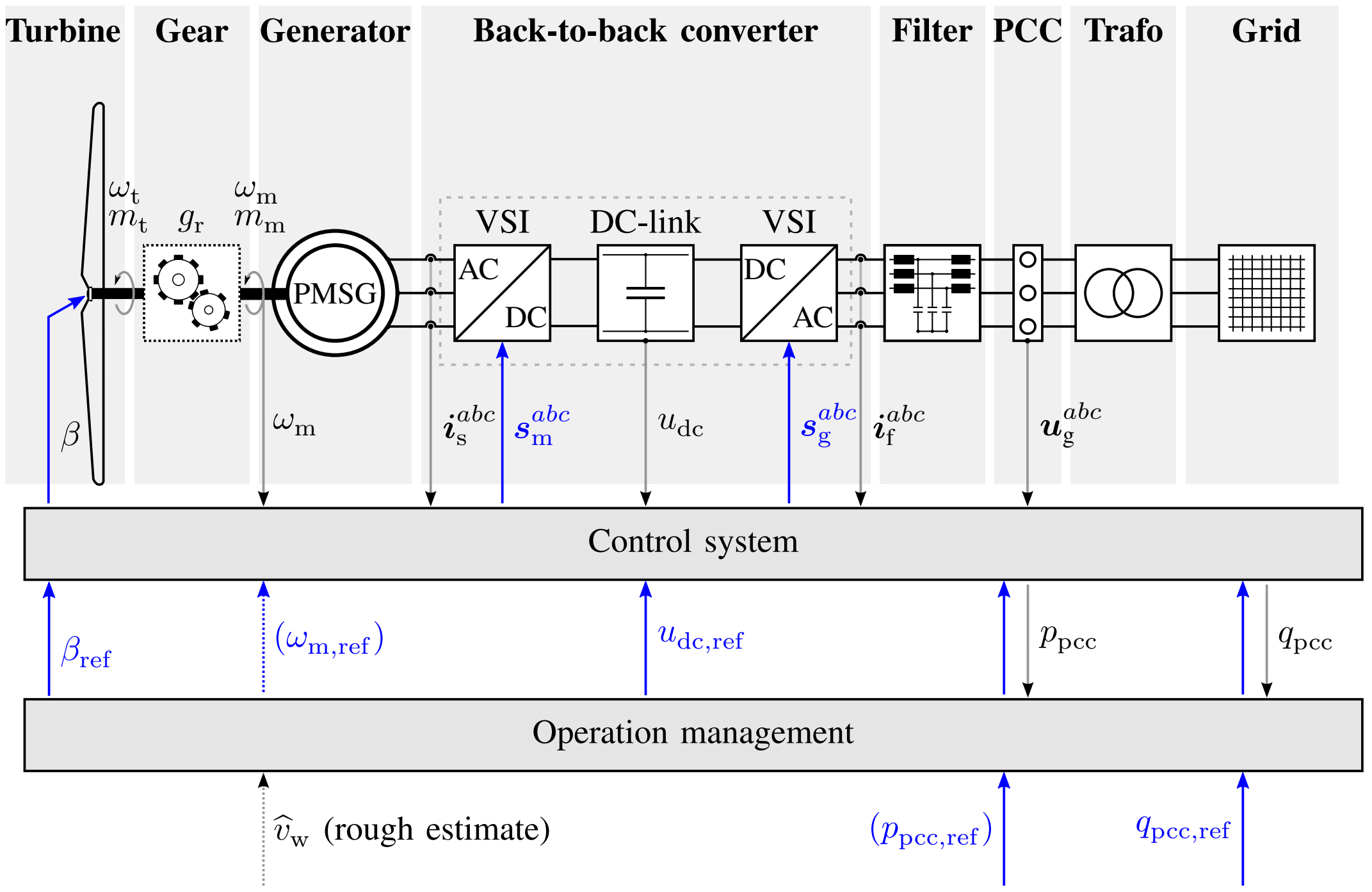

Figure 1: *Overview of the core components of a wind turbine system.*

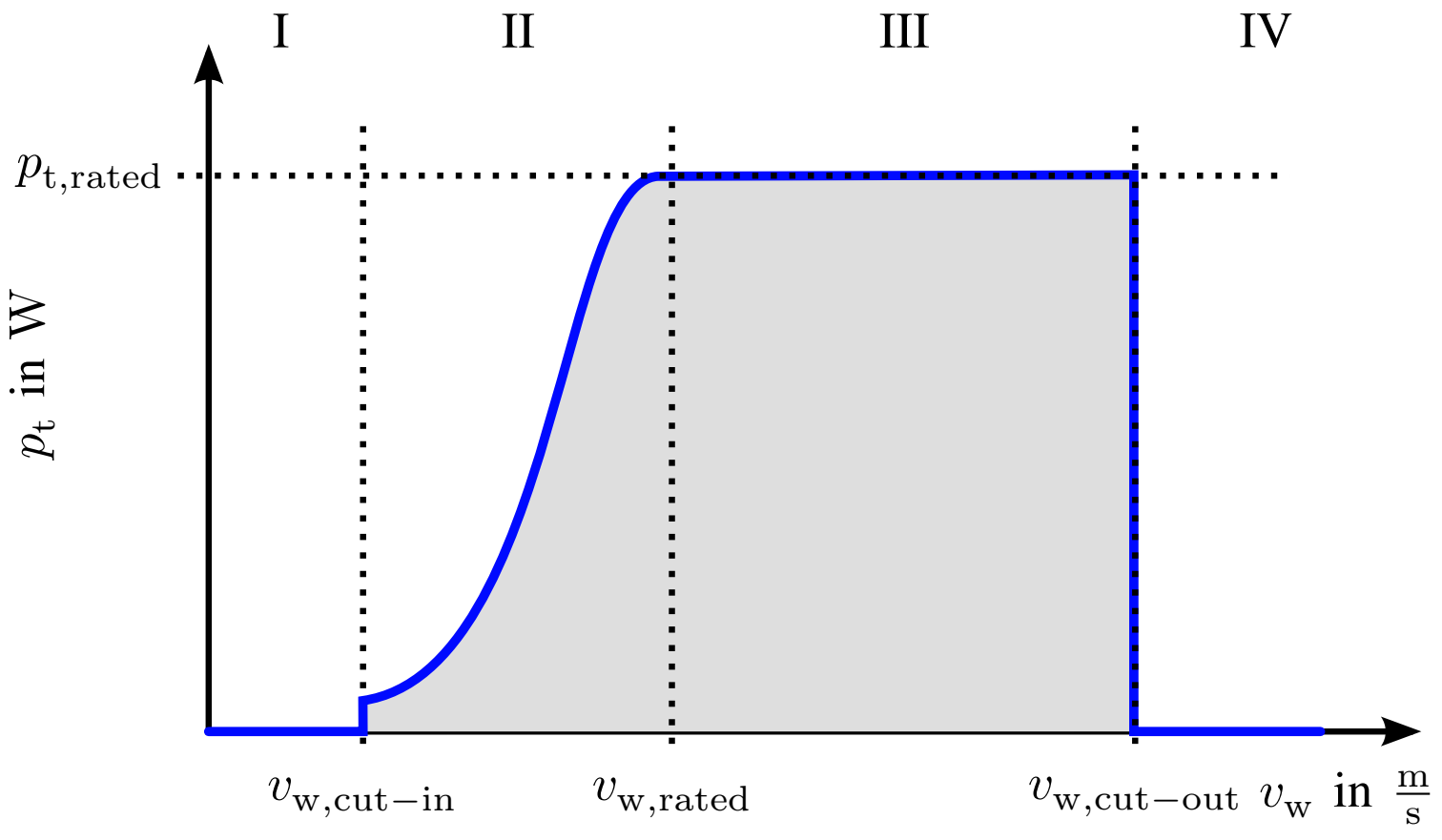

Figure 2: *Operation regimes of a WTS:*

- *Regime I: Standstill (too less wind), i.e.* $p_t = 0$,
- *Regime II: Variable power, i.e.* $0 \leq p_t < p_{t,rated}$ *(Goal: Maximum power point tracking),*
- *Regime III: Nominal power, i.e.* $p_t = p_{t,rated}$,
- *Regime IV: Standstill (too much wind), i.e.* $p_t = 0$.

speed $v_{w,rated}$ (in $\frac{m}{s}$) but at least the (minimum) cut-in wind speed $v_{w,cut-in}$ (in $\frac{m}{s}$). Due to the time-varying nature of the wind speed $v_w(\cdot)$, the turbine output power will vary between zero and nominal power $p_{t,rated}$ (in W), i.e. $0 \leq p_t < p_{t,rated}$. The goal is to extract as much wind power as possible, i.e. *maximum power point tracking* (MPPT) which is achieved by an underlying speed controller (see Sect. IV-A2). In regime III, the wind speed is at least the nominal wind speed but lower than the (maximum) cut-out wind speed $v_{w,cut-out}$ (in $\frac{m}{s}$), i.e. $v_{w,rated} \leq v_w < v_{w,cut-out}$, and the torque $m_m$ of the electrical machine is kept constant at its nominal value (by constant feedforward torque control) and pitch control. The nominal output power is generated, i.e. $p_t = p_{t,rated}$ (see Sect. IV-A3).

In the following sections, the different hardware components illustrated in Fig. 1 (such as turbine, gear, generator, back-to-back converter, filter, PCC) and physical quantities (e.g. $\omega_m$, $\boldsymbol{s}_m^{abc}$, $u_{dc}$) are described and introduced.

## III. Physical Modeling and Control

In this section, the detailed derivation of the complete physical model is presented. It is based on [44] (see also its English translation available at https://arxiv.org/abs/1703.08661) and extended by an explicit representation of the switching behavior and the pitch system dynamics.

### A. Aerodynamics, turbine torque and drive train

*1) Aerodynamics:* The turbine (rotor with three blades) converts part of the kinetic wind energy into rotational energy, which is then converted into electrical energy via the generator. The wind power $p_w(v_w) := \frac{1}{2}\varrho\pi r_t^2 v_w^3$ depends on air density $\varrho$ (in $\frac{\text{kg}}{\text{m}^3}$), rotor radius $r_\text{t}$ (in m), and wind speed $v_\text{w}$ (in $\frac{\text{m}}{\text{s}}$). The extractable turbine power is limited by the Betz limit $c_{\text{p,Betz}} := 16/27$ [46] and is given by

$$p_\text{t}(v_\text{w},\omega_\text{t},\beta) = c_\text{p}(v_\text{w},\omega_\text{t},\beta)\, p_\text{w}(v_\text{w}) \le c_{\text{p,Betz}}\, p_\text{w}(v_\text{w}). \tag{5}$$

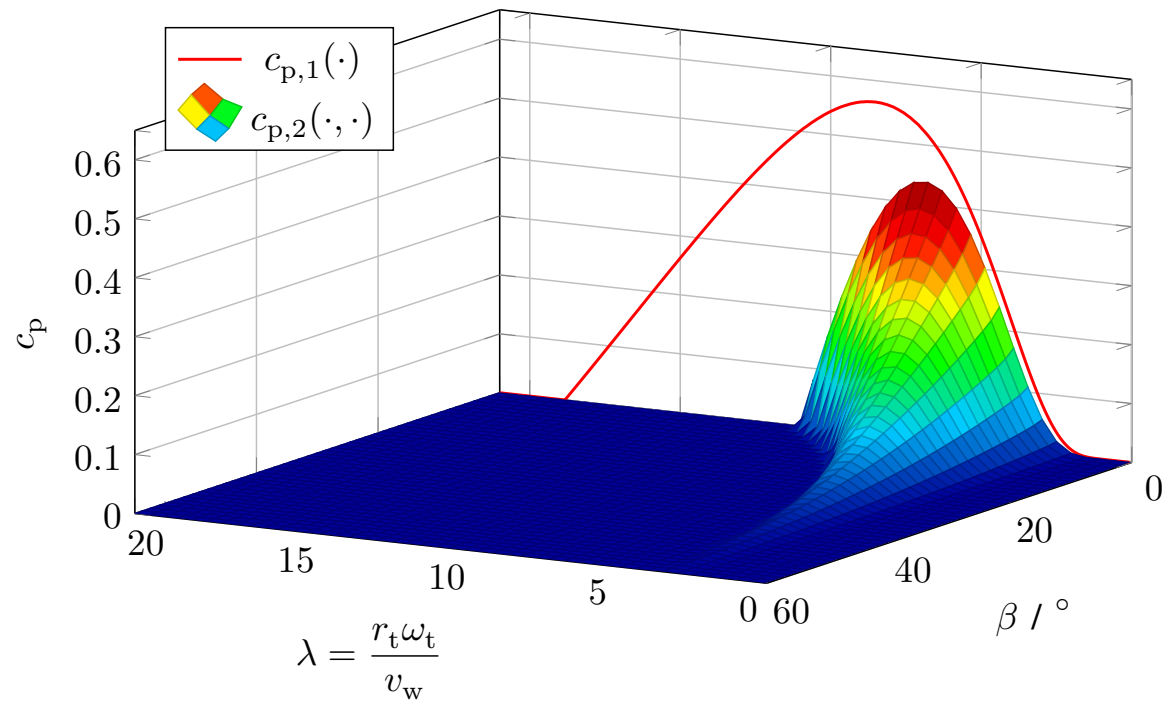

Figure 3: *Graphs of the power coefficient approximations* $c_{\text{p},1}(\cdot)$ *and* $c_{\text{p},2}(\cdot,\cdot)$ *for a* $2\,\text{MW}$ *wind turbine.*

The power coefficient $c_\text{p}(v_\text{w},\omega_\text{t},\beta) = c_\text{p}(\lambda,\beta)$ must be determined for each wind turbine system and is a function of wind speed $v_\text{w}$, turbine angular velocity $\omega_\text{t}$ (in $\frac{\text{rad}}{\text{s}}$), and pitch angle $\beta$ (in °) or of tip speed ratio $\lambda := \lambda(v_\text{w},\omega_\text{t}) := \frac{r_\text{t}\omega_\text{t}}{v_\text{w}}$ (indicating the ratio between the speed at the very end of the rotor blades and the incoming wind speed). Both, tip speed ratio $\lambda$ (or $\omega_\text{t}$) and pitch angle $\beta$, have a direct influence on the amount of power, the wind turbine can extract from the wind. Usually, the power coefficient $c_\text{p}(\lambda,\beta)$ is approximated by the following function [47, (2.38)]

$$\begin{aligned} &c_\text{p}\colon \mathcal{D} \to \mathbb{R}_{\ge 0},\ (\lambda,\beta) \mapsto c_\text{p}(\lambda,\beta) := c_1\big[c_2\, f(\lambda,\beta) - c_3\beta - c_4\beta^k - c_5\big]\, e^{-c_6\, f(\lambda,\beta)} \\ &\text{where } \mathcal{D} := \{\, (\lambda,\beta) \in \mathbb{R}_{>0} \times \mathbb{R}_{\ge 0} \mid c_\text{p}(\lambda,\beta) \ge 0 \,\}. \end{aligned} \tag{6}$$

The constants $c_1,\ldots,c_6 > 0$, the exponent $k \ge 0$ and the continuously differentiable function $f\colon \mathcal{D} \to \mathbb{R}$ can be determined from measurements or by aerodynamic simulations. Two exemplary power coefficient approximations for two different $2\,\text{MW}$ wind turbine systems are as follows [48, Chap. 12]:

- Power coefficient $c_{\text{p},1}(\cdot)$ *without* pitch control system (i.e., $\beta = 0$):

$$c_{\text{p},1}\colon\ \mathcal{D} \to \mathbb{R}_{\ge 0}, \quad (\lambda, 0) \mapsto c_{\text{p},1}(\lambda,0) := c_{\text{p},1}(\lambda) := \Big[46.4 \cdot \big(\tfrac{1}{\lambda} - 0.01\big) - 2.0\Big]\, \text{e}^{-15.6\left(\frac{1}{\lambda} - 0.01\right)}, \tag{7}$$

  which has a global maximum at $\lambda^\star = 8.53$ with $c^\star_{\text{p},1} := c_{\text{p},1}(\lambda^\star) = 0.564$.
- Power coefficient $c_{\text{p},2}(\cdot,\cdot)$ *with* pitch control system (i.e., $\beta \ge 0$):

$$\begin{aligned} c_{\text{p},2}\colon\ \mathcal{D} \to \mathbb{R}_{\ge 0}, \quad (\lambda,\beta) \mapsto \\ c_{\text{p},2}(\lambda,\beta) := 0.73\Big[151\left(\tfrac{1}{\lambda - 0.02\beta} - \tfrac{0.003}{\beta^3+1}\right) - 0.58\beta - 0.002\beta^{2.14} - 13.2\Big]\cdot \\ \cdot \exp\left(-18.4\left(\tfrac{1}{\lambda-0.02\beta} - \tfrac{0.003}{\beta^3+1}\right)\right), \end{aligned} \tag{8}$$

  which, for $\beta = \beta^\star = 0$, has its maximum at $\lambda^\star = 6.91$ with $c^\star_{\text{p},2} := c_{\text{p},2}(\lambda^\star,\beta^\star) = 0.441$.

The graphs of $c_{\text{p},1}(\cdot)$ and $c_{\text{p},2}(\cdot,\cdot)$ are shown in Fig. 3. Both power coefficients are below the possible Betz limit of $c_{\text{p,Betz}} = 16/27 \approx 0.59$. The maximum value of $c_{\text{p},1}(\cdot)$ is larger than that of $c_{\text{p},2}(\cdot,\cdot)$. This does not hold in general but is a characteristic feature of the two wind turbines considered in [49, S. 9], [50], [51].

*2) Pitch control system:* The pitch system allows to control the pitch angle $\beta$ (in °) to its specified reference $\beta_\text{ref}$ (in °). The nonlinear dynamics of the pitch control system are approximated by

$$\tfrac{\text{d}}{\text{d}t}\beta_\diamondsuit(t) = \text{sat}_{-\dot\beta_\text{max}}^{\dot\beta_\text{max}}\left(\tfrac{1}{T_\beta}\big(-\beta(t) + \beta_\text{ref}(t)\big)\right), \qquad \beta_\diamondsuit(0) = \beta_{\diamondsuit,0} \ge 0 \tag{9}$$

$$\beta(t) = \text{sat}_{0^\circ}^{90^\circ}\big(\beta_\diamondsuit(t)\big) \tag{10}$$

with unsaturated pitch angle $\beta_\diamondsuit$ (in °) and its initial value $\beta_{\diamondsuit,0}$ (in °), where $\dot\beta_\text{max} > 0$ (in $\frac{\circ}{\text{s}}$) and $T_\beta$ (in s) are the maximally feasible change rate of the pitch angle and (approximated) pitch control system time constant, respectively (see Fig. 4). The underlying current, speed and position control dynamics are neglected (for details see e.g. [48, Sec. 11.2]). The overall approximated dynamics show the dynamic behavior of a first-order lag system where output and change rate of the state are saturated, respectively (this is a simplification of the pitch dynamics given in [47, Sect. 2.3, 5.5].

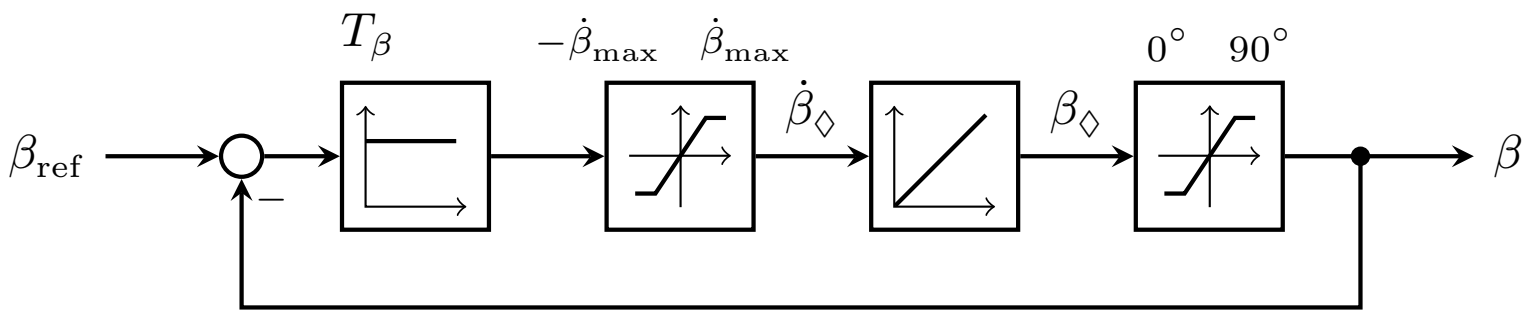

Figure 4: *Block diagram of the approximation of the pitch system dynamics.*

*3) Turbine torque:* The turbine converts the kinetic energy of the wind (translational energy) into rotational energy. Hence, the turbine exerts a torque on the drive train, which leads to an acceleration and rotation of the generator. If friction losses are neglected then, from turbine power $p_{\rm t} = m_{\rm t}\omega_{\rm t}$ with turbine torque $m_{\rm t}$ (in N m) and turbine angular velocity $\omega_{\rm t}$, the turbine torque can directly be computed as follows

$$m_{\rm t}(v_{\rm w}, \lambda, \beta) \overset{(5),(6)}{:=} \tfrac{1}{2}\,\varrho\,\pi\,r_{\rm t}^3\,v_{\rm w}^2\,\frac{c_{\rm p}(\beta,\lambda)}{\lambda} = \tfrac{1}{2}\,\varrho\,\pi\,r_{\rm t}^2\,v_{\rm w}^3\,\frac{c_{\rm p}(v_{\rm w},\omega_{\rm t},\beta)}{\omega_{\rm t}} =: m_{\rm t}(v_{\rm w}, \omega_{\rm t}, \beta). \tag{11}$$

The turbine torque is a *nonlinear function* of pitch angle $\beta$, wind speed $v_{\rm w}$ and tip speed ratio $\lambda$ or turbine angular velocity $\omega_{\rm t}$.

**Remark III.1.** *The approximation* (6) *of the power coefficient* $c_{\rm p}(\cdot,\cdot)$ *does* not *allow for the simulation of the start-up of a wind turbine system, since* $\lim_{\omega_{\rm t}\to 0} m_{\rm t}(v_{\rm w},\omega_{\rm t},\beta) = 0$ *for all* $v_{\rm w} > 0$ *and* $\beta \geq 0$ *[44]. The approximation* (6) *only yields physically meaningful results for* $\lambda > 0$. *At standstill, the accelerating torque would be zero.*

*4) Drive train:* A gearbox transmits the mechanical turbine power via a shaft to the rotor of the generator. In modern wind turbine system, the turbine angular velocity $\omega_{\rm t}$ is significantly lower than the angular velocity $\omega_{\rm m}$ (in $\frac{\rm rad}{\rm s}$) of the machine (generator). Therefore, a step-up gearbox with ratio $g_{\rm r} \gg 1$ is usually employed (exceptions are wind turbines with "Direct Drive", i.e. $g_{\rm r} = 1$, where the generator is connected directly to the turbine rotor). For a rigid coupling, generator (machine) and turbine angular velocities are related by $\omega_{\rm m} = g_{\rm r}\omega_{\rm t}$. Hence, denoting the machine torque by $m_{\rm m}$ (in N m), turbine power $p_{\rm t}$ and turbine torque $m_{\rm t}$ are converted to the machine-side quantities as follows

$$p_{\rm m} = \omega_{\rm m} m_{\rm m} = g_{\rm r}\omega_{\rm t}\frac{m_{\rm t}}{g_{\rm r}} = p_{\rm t}. \tag{12}$$

Moreover, the inertias $\Theta_{\rm t}$ and $\Theta_{\rm m}$ (both in kg m$^2$) of turbine rotor (+hub) and machine rotor (+shaft) can be merged to the overall inertia $\Theta := \frac{\Theta_{\rm t}}{g_{\rm r}^2} + \Theta_{\rm m}$ of the drive train. For simplicity, turbine-side and machine-side friction and elasticity in the shaft are neglected (for more details on friction modeling & compensation, and elastic drive train modeling, see [48, Sect. 11.1.5 & Chap. 12] and [52]).

### *B. Electrical system in three-phase* $(a, b, c)$*-reference frame*

In Fig. 5, the (simplified) electrical network of a wind turbine system with permanent-magnet synchronous or induction generator is depicted. The machine-side network (left) shows the stator windings with stator phase voltages $\boldsymbol{u}_{\rm s}^{abc} = (u_{\rm s}^a, u_{\rm s}^b, u_{\rm s}^c)^\top$ (each in V), stator phase current $\boldsymbol{i}_{\rm s}^{abc} = (i_{\rm s}^a, i_{\rm s}^b, i_{\rm s}^c)^\top$ (each in A), stator phase resistance $R_{\rm s}$ (in Ω), and stator flux linkage $\boldsymbol{\psi}_{\rm s}^{abc} = (\psi_{\rm s}^a, \psi_{\rm s}^b, \psi_{\rm s}^c)^\top$ (each in Wb). The grid-side network (right) comprises filter and grid with filter phase voltages $\boldsymbol{u}_{\rm f}^{abc} = (u_{\rm f}^a, u_{\rm f}^b, u_{\rm f}^c)^\top$ (each in V), filter phase currents $\boldsymbol{i}_{\rm f}^{abc} = (i_{\rm f}^a, i_{\rm f}^b, i_{\rm f}^c)^\top$ (each in A), filter resistance $R_{\rm f}$ (in Ω), filter inductance $L_{\rm f}$ (in H) and the grid phase voltages $\boldsymbol{u}_{\rm g}^{abc} = (u_{\rm g}^a, u_{\rm g}^b, u_{\rm g}^c)^\top$ (each in V). The (stepped-down) grid voltage $\boldsymbol{u}_{\rm g}^{abc}$ is measured at the *Point of Common Coupling (PCC)*. The transmission ratio of the transformer (not shown in Fig. 5) is not explicitly modeled. The back-to-back converter electrically links machine and grid side. It consists of two fully-controlled voltage source converters (VSCs) which share a common DC-link with voltage $u_{\rm dc}$ (in V). Machine-side and grid-side converters exchange the stator power $p_{\rm s}$ (in W) and the filter power $p_{\rm f}$ (in W) via the DC-link. In the continuous operation of the wind turbine, the DC-link capacitance $C_{\rm dc}$ (in F) on average is *not* charged or discharged and no DC-link power, $p_{\rm dc}$ (in W), is exchanged within the circuit and the DC-link voltage $u_{\rm dc}$ remains (almost) constant. The (active) power $p_{\rm pcc}$ (in W) is fed into the grid at the PCC.

*1) Machine-side dynamics (electrical machine/generator and drive train):* For an isotropic PMSG Kirchhoff's laws (see the electrical circuit in Fig. 5) and Newton mechanics yield the following fifth-order dynamic system (for details see [48,

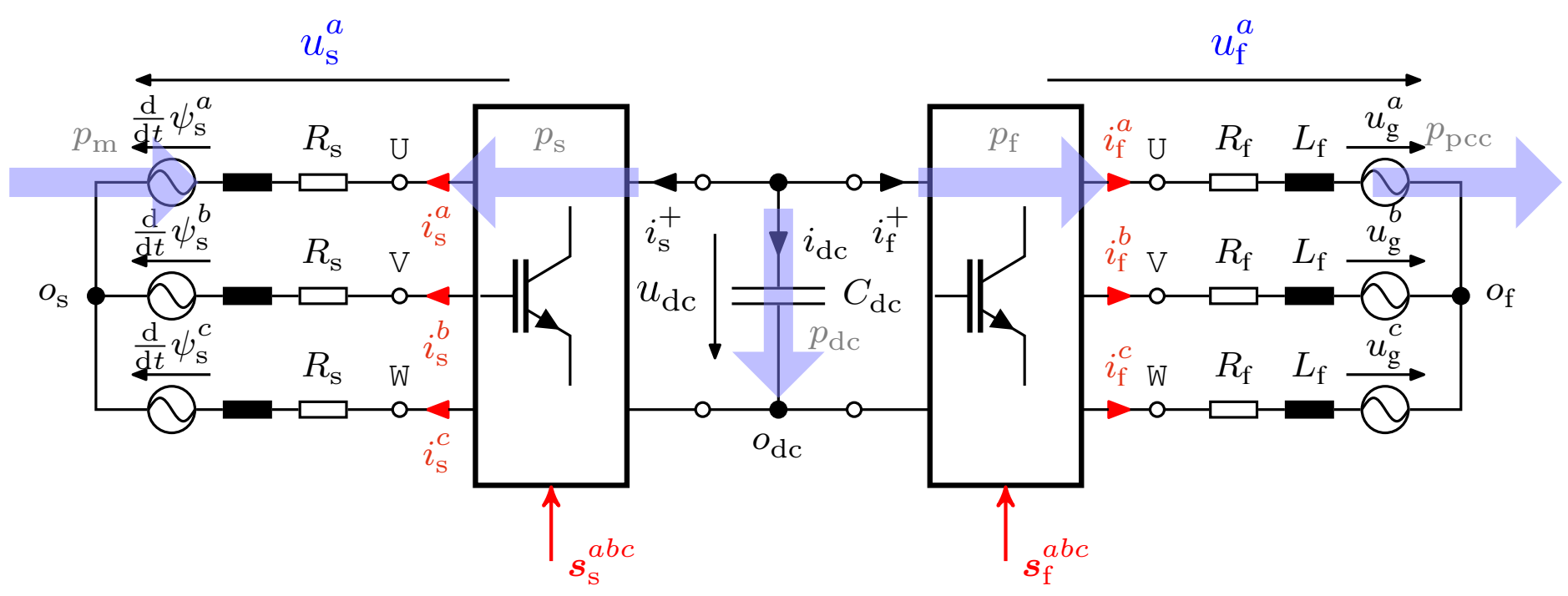

Figure 5: *Electrical network of overall wind turbine system: Permanent-magnet synchronous generator (left), back-to-back converter (middle) sharing a common DC-link, grid-side filter, point of common coupling (PCC) and balanced grid (right, neglecting the transformer).*

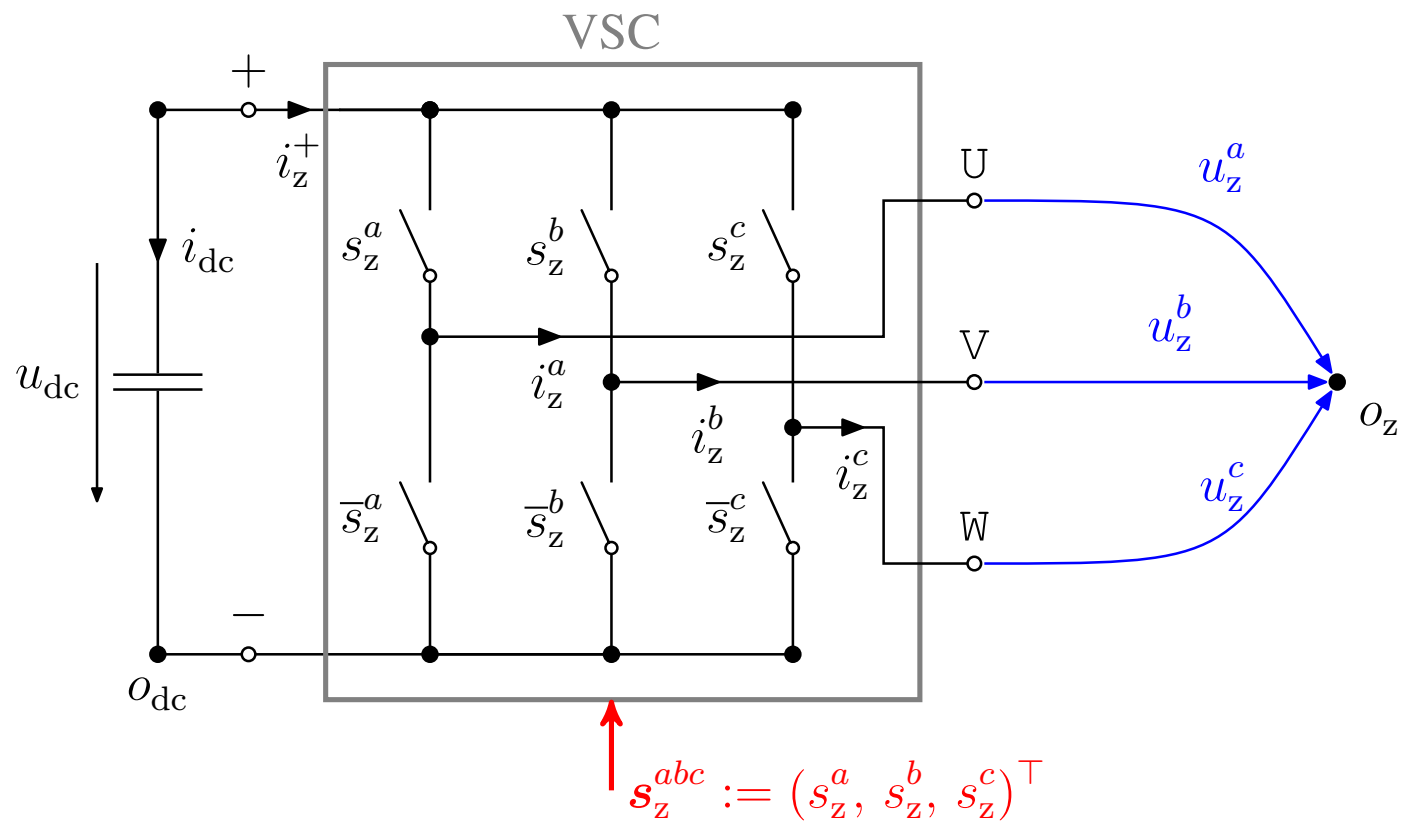

Figure 6: *Electrical circuit of voltage source converter (VSC) with ideal switches and DC-link ($\mathrm{z} \in \{\mathrm{s}, \mathrm{f}\}$).*

Chap. 14])

$$\left.\begin{aligned}
\tfrac{\mathrm{d}}{\mathrm{d}t}\boldsymbol{i}_\mathrm{s}^{abc}(t) &= \left(\boldsymbol{L}_\mathrm{s}^{abc}\right)^{-1}\left[\boldsymbol{u}_\mathrm{s}^{abc}(t) - R_\mathrm{s}\boldsymbol{i}_\mathrm{s}^{abc}(t) + n_\mathrm{p}\omega_\mathrm{m}(t)\widehat{\psi}_\mathrm{pm}\begin{pmatrix}\sin(n_\mathrm{p}\phi_\mathrm{m}(t))\\ \sin(n_\mathrm{p}\phi_\mathrm{m}(t) - \frac{2}{3}\pi)\\ \sin(n_\mathrm{p}\phi_\mathrm{m}(t) - \frac{4}{3}\pi)\end{pmatrix}\right] &&, \boldsymbol{i}_\mathrm{s}^{abc}(0) = \boldsymbol{i}_{\mathrm{s},0}^{abc}\\
\tfrac{\mathrm{d}}{\mathrm{d}t}\omega_\mathrm{m}(t) &= \tfrac{1}{\Theta}\Bigg[\frac{m_\mathrm{t}\left(v_\mathrm{w}(t), \omega_\mathrm{m}(t)/g_\mathrm{r}, \beta(t)\right)}{g_\mathrm{r}} + \underbrace{n_\mathrm{p}\boldsymbol{i}_\mathrm{s}^{abc}(t)^\top \boldsymbol{J}_\Sigma \boldsymbol{\psi}_\mathrm{pm}^{abc}(\phi_\mathrm{m}(t))}_{=: m_\mathrm{m}\left(\boldsymbol{i}_\mathrm{s}^{abc}(t), \phi_\mathrm{m}(t)\right)}\Bigg] &&, \omega_\mathrm{m}(0) = \omega_{\mathrm{m},0}\\
\tfrac{\mathrm{d}}{\mathrm{d}t}\phi_\mathrm{m}(t) &= \omega_\mathrm{m}(t) &&, \phi_\mathrm{m}(0) = \phi_{\mathrm{m},0}
\end{aligned}\right\} \tag{13}$$

where

$$\boldsymbol{L}_\mathrm{s}^{abc} := \begin{bmatrix} L_{\mathrm{s,m}} + L_{\mathrm{s},\sigma} & -\frac{L_{\mathrm{s,m}}}{2} & -\frac{L_{\mathrm{s,m}}}{2}\\ -\frac{L_{\mathrm{s,m}}}{2} & L_{\mathrm{s,m}} + L_{\mathrm{s},\sigma} & -\frac{L_{\mathrm{s,m}}}{2}\\ -\frac{L_{\mathrm{s,m}}}{2} & -\frac{L_{\mathrm{s,m}}}{2} & L_{\mathrm{s,m}} + L_{\mathrm{s},\sigma}\end{bmatrix} = \left(\boldsymbol{L}_\mathrm{s}^{abc}\right)^\top > 0 \quad \text{and} \quad \boldsymbol{\psi}_\mathrm{pm}^{abc}(\phi_\mathrm{m}) := \widehat{\psi}_\mathrm{pm}\begin{pmatrix}\cos(n_\mathrm{p}\phi_\mathrm{m})\\ \cos(n_\mathrm{p}\phi_\mathrm{m} - \frac{2}{3}\pi)\\ \cos(n_\mathrm{p}\phi_\mathrm{m} - \frac{4}{3}\pi)\end{pmatrix} \tag{14}$$

are inductance matrix $\boldsymbol{L}_\mathrm{s}^{abc}$ (with mutual inductance $L_{\mathrm{s,m}}$ and leakage inductance $L_{\mathrm{s},\sigma}$ such that $\boldsymbol{L}_\mathrm{s}^{abc} > 0$; both in H) and permanent-magnet flux linkage vector $\boldsymbol{\psi}_\mathrm{pm}^{abc}$ (each in Wb), respectively; $\boldsymbol{J}_\Sigma$ is as in (4). The dynamics in (13) incorporate electrical stator dynamics (the first three states) with stator flux linkage $\boldsymbol{\psi}_\mathrm{s}^{abc}(\boldsymbol{i}_\mathrm{s}^{abc}, \phi_\mathrm{m}) = \boldsymbol{L}_\mathrm{s}^{abc}\boldsymbol{i}_\mathrm{s}^{abc} + \boldsymbol{\psi}_\mathrm{pm}^{abc}(\phi_\mathrm{m})$ and the rotatory (mechanical) dynamics of the generator. The stator currents $\boldsymbol{i}_\mathrm{s}^{abc}$ and the mechanical angular velocity $\omega_\mathrm{m}$ (and/or the mechanical angle $\phi_\mathrm{m}$) are measured and available for feedback control.

*2) Power electronics and DC-link dynamics (back-to-back converter):* Although multi-level converters for the regulation of wind power plants are likely to be used in the future [53], the still widely used two-level back-to-back converter will be considered in this section. Each of the two voltage source converters (VSCs) of the back-to-back converter can be modeled as illustrated in Fig. 6 by ideal switches (dynamics of the semi-conductors and free-wheeling diodes are neglected). The stator-side converter feeds the electrical machine (generator) with the phase voltages $\boldsymbol{u}_\mathrm{s}^{abc}$, while the filter-side converter applies the phase voltages $\boldsymbol{u}_\mathrm{f}^{abc}$ to the line filter. The output voltages of the converters depend on the DC-link voltage $u_\mathrm{dc}$ and are generated by adequate modulation, i.e. the application of an adequate sequence of the switching vectors $\boldsymbol{s}_\mathrm{s}^{abc} = (s_\mathrm{s}^a,\, s_\mathrm{s}^b,\, s_\mathrm{s}^c)^\top$ and $\boldsymbol{s}_\mathrm{f}^{abc} = (s_\mathrm{f}^a,\, s_\mathrm{f}^b,\, s_\mathrm{f}^c)^\top$. For the considered two-level voltage source converters, there exist eight possible switching vectors $(\boldsymbol{s}_\mathrm{z}^{abc})^\top \in \mathbb{S}_8 := \{\mathtt{000}, \mathtt{100}, \ldots, \mathtt{111}\}$ for $\mathrm{z} \in \{\mathrm{s}, \mathrm{f}\}$ (see Fig. 6). Moreover, for balanced[2] voltages, the stator voltages and filter voltages are given by

$$\forall \mathrm{z} \in \{\mathrm{s}, \mathrm{f}\}\colon\ \boldsymbol{u}_\mathrm{z}^{abc}(\boldsymbol{s}_\mathrm{z}^{abc}, u_\mathrm{dc}) = u_\mathrm{dc} \underbrace{\tfrac{1}{3}\begin{bmatrix} 2 & -1 & -1 \\ -1 & 2 & -1 \\ -1 & -1 & 2 \end{bmatrix}}_{=:\boldsymbol{T}_\mathrm{vsc}} \boldsymbol{s}_\mathrm{z}^{abc} \implies \|\boldsymbol{u}_\mathrm{z}^{abc}(\boldsymbol{s}_\mathrm{z}^{abc}, u_\mathrm{dc})\| \leq \widehat{u} := \tfrac{2}{3} u_\mathrm{dc}. \tag{15}$$

Due to the limited DC-link voltage, each VSC can generate only a constrained phase voltage amplitude. The shared DC-link capacitor $C_\mathrm{dc}$ is charged or discharged via the DC-link current $i_\mathrm{dc} = -i_\mathrm{m}^{+} - i_\mathrm{f}^{+}$ (in A) which depends on the machine-side and grid-side currents, respectively (see Fig. 5). The DC-link dynamics are given by

$$\tfrac{\mathrm{d}}{\mathrm{d}t} u_\mathrm{dc}(t) = \tfrac{1}{C_\mathrm{dc}} i_\mathrm{dc}(t) = \tfrac{1}{C_\mathrm{dc}}\Big( - \underbrace{\boldsymbol{i}_\mathrm{s}^{abc}(t)^\top \boldsymbol{s}_\mathrm{s}^{abc}(t)}_{=:i_\mathrm{s}^{+}(t)} - \underbrace{\boldsymbol{i}_\mathrm{f}^{abc}(t)^\top \boldsymbol{s}_\mathrm{f}^{abc}(t)}_{=:i_\mathrm{f}^{+}(t)} \Big), \qquad u_\mathrm{dc}(0) = u_\mathrm{dc}^0 > 0\,(\text{in V}). \tag{16}$$

The state-of-the-art modulation technique is the *space vector modulation* (SVM) which can reproduce *average phase voltage amplitudes* up to $\bar{u} := u_\mathrm{dc}/\sqrt{3}$. The classical *pulse width modulation* (PWM; without over-modulation) can reproduce average phase voltage amplitudes up to $\bar{u} := u_\mathrm{dc}/2$ (see [54, S. 658–720] and [55, S. 132–136]). Consider a feasible stator reference phase voltage vector $\boldsymbol{u}_\mathrm{s,ref}^{abc}(\cdot) \in \mathcal{C}(\mathbb{R}_{\geq 0}; [-\bar{u}, \bar{u}]^3)$ and a carrier signal $c_\wedge(\cdot) \in \mathcal{C}(\mathbb{R}_{\geq 0}; [-1, 1])$ (e.g. a sawtooth or triangular carrier signal with period $T_\mathrm{sw} = 1/f_\mathrm{sw}$ (in s) inversely proportional to the switching frequency $f_\mathrm{sw}$ (in Hz)). Then, PWM generates its pulse pattern by a simple and instantaneous comparison of normalized reference phase voltages and carrier signal. More precisely, the actual switching signal vector for PWM is obtained by

$$\boldsymbol{s}_\mathrm{z}^{abc}\big(\boldsymbol{u}_\mathrm{z,ref}^{abc}, u_\mathrm{dc}, t\big) := \begin{pmatrix} \sigma\Big(\frac{u_\mathrm{z,ref}^{a}}{u_\mathrm{dc}/2} - c_\wedge(t)\Big) \\ \sigma\Big(\frac{u_\mathrm{z,ref}^{b}}{u_\mathrm{dc}/2} - c_\wedge(t)\Big) \\ \sigma\Big(\frac{u_\mathrm{z,ref}^{c}}{u_\mathrm{dc}/2} - c_\wedge(t)\Big) \end{pmatrix} =: \boldsymbol{\sigma}\Big(\frac{\boldsymbol{u}_\mathrm{z,ref}^{abc}}{u_\mathrm{dc}/2} - \boldsymbol{1}_3 c_\wedge(t)\Big) \in \mathbb{S}_8 \tag{17}$$

where $\sigma(\cdot)$ is the Heaviside (step) function defined by $\sigma\colon \mathbb{R} \to \{0, 1\}$, $\gamma \mapsto \sigma(\gamma) := \begin{cases} 1 & ,\gamma \geq 0 \\ 0 & ,\gamma < 0 \end{cases}$. For SVM, the reference voltage vector $\boldsymbol{u}_\mathrm{z,ref}^{abc} = (u_\mathrm{z,ref}^{a}, u_\mathrm{z,ref}^{b}, u_\mathrm{z,ref}^{c})^\top$ in (17) must be replaced [56, p. 267–271] by the following expression $\boldsymbol{u}_\mathrm{z,SVM,ref}^{abc} = \Big[\boldsymbol{u}_\mathrm{z,ref}^{abc} - \frac{\mathbf{max}(\boldsymbol{u}_\mathrm{z,ref}^{abc}) + \mathbf{min}(\boldsymbol{u}_\mathrm{z,ref}^{abc})}{2}\Big]$ where $\boldsymbol{u}_\mathrm{z,ref}^{abc}$ is the (original) reference vector from the control system and $\mathbf{max}(\boldsymbol{\xi}) := \max(\boldsymbol{\xi})\boldsymbol{1}_3$ and $\mathbf{min}(\boldsymbol{\xi}) := \min(\boldsymbol{\xi})\boldsymbol{1}_3$ return the minimal and maximal entries of the vector $\boldsymbol{\xi} \in \mathbb{R}^3$, respectively. The reference voltages are normalized with respect to $u_\mathrm{dc}/2$. For each phase $p \in \{a, b, c\}$ and $\mathrm{z} \in \{\mathrm{s}, \mathrm{f}\}$, the phase switching signal is high, i.e. $s_\mathrm{z}^{p}(t) = 1$, when the normalized reference phase voltage is larger than or equal to the carrier signal, i.e. $u_\mathrm{z,ref}^{p}(t) \geq c_\wedge(t)$; whereas the switching signal is low, i.e. $s_\mathrm{z}^{p}(t) = 0$, when the normalized reference is smaller than the carrier signal, i.e. $u_\mathrm{z,ref}^{p}(t) < c_\wedge(t)$.

Due to the finite (eight) number of switching vectors, not all reference voltage vectors can be generated instantaneously. The converter exhibits some delay which is inversely proportional to the switching frequency $f_\mathrm{sw} \gg 1\,\mathrm{Hz}$ [57, p. 525-526]. On average, this delay can be quantified by the *inverter delay time* $T_\mathrm{avg}$ (in s) which is required to produce the *average output phase voltage vector* defined by [48, Chapter 14]

$$\forall\, t \geq T_\mathrm{avg}\colon \quad \overline{\boldsymbol{u}}_\mathrm{z}^{abc}(t) := \tfrac{1}{T_\mathrm{avg}} \int_{t - T_\mathrm{avg}}^{t} \boldsymbol{u}_\mathrm{z}^{abc}(\tau)\,\mathrm{d}\tau \approx \boldsymbol{u}_\mathrm{z,ref}^{abc}(t - T_\mathrm{avg}). \tag{18}$$

The delay varies within the interval $T_\mathrm{avg} \in \big[\frac{1}{2 f_\mathrm{sw}}, \frac{3}{2 f_\mathrm{sw}}\big]$ [58] and depends on switching frequency $f_\mathrm{sw}$, the selected modulation scheme (e.g. PWM or SVM) and its implementation[3] (e.g. on FPGA, DSP or micro-processor).

*3) Grid-side dynamics (filter, Point of Common Coupling (PCC) and grid):* To induce sinusoidal phase currents to the power grid, a line filter must be used to filter out the switching behavior of the VSC. A simple $RL$-filter (in each phase) with filter inductance $L_\mathrm{f}$ (in H) and filter resistance $R_\mathrm{f}$ (in $\Omega$) is considered. The grid-side converter generates the (filter) voltages $\boldsymbol{u}_\mathrm{f}^{abc}$ which are applied to the filter and, due to the inductance $L_\mathrm{f}$, lead to (approximately) sinusoidal filter phase currents $\boldsymbol{i}_\mathrm{f}^{abc}$. In the filter resistance $R_\mathrm{f}$, the copper losses $R_\mathrm{f}\|\boldsymbol{i}_\mathrm{f}^{abc}\|^2$ (in W) are dissipated and converted into heat. The grid-side electrical network with grid voltages $\boldsymbol{u}_\mathrm{g}^{abc}$ is shown in Fig. 5. According to Kirchoff's voltage law, the grid-side dynamics are given by

$$\tfrac{\mathrm{d}}{\mathrm{d}t} \boldsymbol{i}_\mathrm{f}^{abc}(t) = \tfrac{1}{L_\mathrm{f}} \Bigg[ \boldsymbol{u}_\mathrm{f}^{abc}(t) - R_\mathrm{f} \boldsymbol{i}_\mathrm{f}^{abc}(t) - \underbrace{\hat{u}_\mathrm{g}(t) \begin{pmatrix} \cos(\phi_\mathrm{g}(t)) \\ \cos(\phi_\mathrm{g}(t) - \frac{2\pi}{3}) \\ \cos(\phi_\mathrm{g}(t) - \frac{4\pi}{3}) \end{pmatrix}}_{=:\boldsymbol{u}_\mathrm{g}^{abc}(t)} \Bigg], \quad \boldsymbol{i}_\mathrm{f}^{abc}(0) = \boldsymbol{i}_\mathrm{f,0}^{abc} \in \mathbb{R}^3, \tag{19}$$

[2] I.e. the following holds $u_\mathrm{s}^{a}(t) + u_\mathrm{s}^{b}(t) + u_\mathrm{s}^{c}(t) = u_\mathrm{f}^{a}(t) + u_\mathrm{f}^{b}(t) + u_\mathrm{f}^{c}(t) = 0$ for all $t \geq 0$.

[3] For most modern implementations, the reference voltages are sampled with the switching frequency $f_\mathrm{sw}$ at the maximum (or minimum) of the carrier signal $c_\wedge(\cdot)$ and hold constant over the period $T_\mathrm{sw} = 1/f_\mathrm{sw}$, i.e. "*symmetrical sampling*" [48, Chapter 14], [56, Chapter 3.6].

where the grid voltages $\boldsymbol{u}_{\mathrm{g}}^{abc}(\cdot)$ depend on a (possibly time-varying) amplitude $\hat{u}_{\mathrm{g}}(\cdot) \geq 0\,\mathrm{V}$ and on a time-varying grid angle $\phi_{\mathrm{g}}(\cdot) := \int_0^{\cdot} \omega_{\mathrm{g}}(\tau)\,\mathrm{d}\tau + \phi_{\mathrm{g},0}$ (in rad). The angular grid frequency[4] $\omega_{\mathrm{g}}$ (in $\frac{\mathrm{rad}}{\mathrm{s}}$) might also vary over time and the initial phase angle $\phi_{\mathrm{g},0}$ is usually unknown (a phase-locked loop is employed to detect $\phi_{\mathrm{g},0}$ and $\omega_{\mathrm{g}}$ [44]). The grid voltages $\boldsymbol{u}_{\mathrm{g}}^{abc}$ are measured before (or after) the transformer. The transformer steps up the voltage to a higher voltage level at the PCC (for example, to the medium voltage level of the power grid).

**Remark III.2.** *In wind turbine systems, also $LCL$-filters are used instead of $RL$-filters. The design of an $LCL$-filter allows for smaller inductances. Thus, an $LCL$-filter can be made smaller than an $RL$-filter. A detailed discussion of the $LCL$-filter design can be found in [1, Kap. 11]. The control of grid-side power converters connected to the grid via $LCL$-filters is discussed in e.g. [60], [61].*

*4) Power output:* The WECS outputs the active and reactive instantaneous powers $p_{\mathrm{pcc}}$ (in W) and $q_{\mathrm{pcc}}$ (in var) at the point of common coupling (pcc) which, for $\boldsymbol{J}_{\Sigma}$ as in (4) and $\boldsymbol{u}_{\mathrm{g}}^{abc}(t)$ as in (19), are respectively given by

$$p_{\mathrm{pcc}}(\boldsymbol{i}_{\mathrm{f}}^{abc}, t) = \boldsymbol{u}_{\mathrm{g}}^{abc}(t)^{\top} \boldsymbol{i}_{\mathrm{f}}^{abc} \qquad \text{and} \qquad q_{\mathrm{pcc}}(\boldsymbol{i}_{\mathrm{f}}^{abc}, t) = \boldsymbol{u}_{\mathrm{g}}^{abc}(t)^{\top} \boldsymbol{J}_{\Sigma} \boldsymbol{i}_{\mathrm{f}}^{abc}. \tag{20}$$

*5) Overall dynamics in nonlinear state-space representation:* The overall model is of eleven-th order and considers switching. For state vector

$$\begin{aligned}\boldsymbol{x} &:= \Big( \underbrace{(x_1,\, x_2,\, x_3)}_{=:\boldsymbol{x}_{1\text{-}3}^{\top}},\quad x_4,\quad x_5,\quad x_6,\quad \underbrace{(x_7,\, x_8,\, x_9)}_{=:\boldsymbol{x}_{7\text{-}9}^{\top}},\quad x_{10},\quad x_{11} \Big)^{\top} \\ &:= \Big( (\boldsymbol{i}_{\mathrm{s}}^{abc})^{\top},\quad \omega_{\mathrm{m}},\quad \phi_{\mathrm{m}},\quad u_{\mathrm{dc}},\quad (\boldsymbol{i}_{\mathrm{f}}^{abc})^{\top},\quad \phi_{\mathrm{g}},\quad \beta_{\diamondsuit} \Big)^{\top} \in \mathbb{R}^{11},\end{aligned}$$

and control input vector

$$\boldsymbol{u} := \Big( \underbrace{(u_1,\, u_2,\, u_3)}_{=:\boldsymbol{u}_{1\text{-}3}^{\top}},\quad \underbrace{(u_4,\, u_5,\, u_6)}_{=:\boldsymbol{u}_{4\text{-}6}^{\top}},\quad u_7 \Big)^{\top} := \Big( (\boldsymbol{u}_{\mathrm{s,ref}}^{abc})^{\top},\quad (\boldsymbol{u}_{\mathrm{f,ref}}^{abc})^{\top}, \beta_{\mathrm{ref}} \Big)^{\top} \in \mathbb{R}^{7},$$

the overall system dynamics with output $\boldsymbol{y}$ are given by the following nonlinear ordinary differential equation

$$\left.\begin{aligned}
\frac{\mathrm{d}}{\mathrm{d}t}\boldsymbol{x} &= \underbrace{\begin{pmatrix}
(\boldsymbol{L}_{\mathrm{s}}^{abc})^{-1} \left[ x_6 \boldsymbol{T}_{\mathrm{vsc}} \boldsymbol{\sigma}\big(\frac{\boldsymbol{u}_{1\text{-}3}}{x_6/2} - \boldsymbol{1}_3 c_{\wedge}(t)\big) - R_{\mathrm{s}} \boldsymbol{x}_{1\text{-}3} + n_{\mathrm{p}} x_4 \widehat{\psi}_{\mathrm{pm}} \begin{pmatrix} \sin(n_{\mathrm{p}} x_5) \\ \sin(n_{\mathrm{p}} x_5 - \frac{2}{3}\pi) \\ \sin(n_{\mathrm{p}} x_5 - \frac{4}{3}\pi) \end{pmatrix} \right] \\
\frac{1}{\Theta} \left[ \varrho\, \pi\, r_{\mathrm{t}}^2\, v_{\mathrm{w}}(t)^3 \frac{c_{\mathrm{p}}\big(r_{\mathrm{t}} x_4 / (g_{\mathrm{r}} v_{\mathrm{w}}(t)),\, \mathrm{sat}_{0^\circ}^{90^\circ}(x_{11})\big)}{2 x_4} + n_{\mathrm{p}} \widehat{\psi}_{\mathrm{pm}} \boldsymbol{x}_{1\text{-}3}^{\top} \boldsymbol{J}_{\Sigma} \begin{pmatrix} \cos(n_{\mathrm{p}} x_5) \\ \cos(n_{\mathrm{p}} x_5 - \frac{2}{3}\pi) \\ \cos(n_{\mathrm{p}} x_5 - \frac{4}{3}\pi) \end{pmatrix} \right] \\
x_4 \\
\frac{1}{C_{\mathrm{dc}}} \left[ - \boldsymbol{x}_{1\text{-}3}^{\top} \boldsymbol{\sigma}\big(\frac{\boldsymbol{u}_{1\text{-}3}}{x_6/2} - \boldsymbol{1}_3 c_{\wedge}(t)\big) - \boldsymbol{x}_{7\text{-}9}^{\top} \boldsymbol{\sigma}\big(\frac{\boldsymbol{u}_{4\text{-}6}}{x_6/2} - \boldsymbol{1}_3 c_{\wedge}(t)\big) \right] \\
\frac{1}{L_{\mathrm{f}}} \left[ x_6 \boldsymbol{T}_{\mathrm{vsc}} \boldsymbol{\sigma}\big(\frac{\boldsymbol{u}_{4\text{-}6}}{x_6/2} - \boldsymbol{1}_3 c_{\wedge}(t)\big) - R_{\mathrm{f}} \boldsymbol{x}_{7\text{-}9} - \hat{u}_{\mathrm{g}}(t) \begin{pmatrix} \cos(x_{10}) \\ \cos(x_{10} - \frac{2\pi}{3}) \\ \cos(x_{10} - \frac{4\pi}{3}) \end{pmatrix} \right] \\
\omega_{\mathrm{g}}(t) \\
\mathrm{sat}_{-\dot{\beta}_{\max}}^{\dot{\beta}_{\max}} \left( \frac{1}{T_{\beta}} \big( - \mathrm{sat}_{0^\circ}^{90^\circ}(x_{11}) + u_7 \big) \right)
\end{pmatrix}}_{=:\boldsymbol{f}_{\mathrm{abc}}(\boldsymbol{x}, \boldsymbol{u}, t) \in \mathbb{R}^{11}} \\
\boldsymbol{y} &= \underbrace{\hat{u}_{\mathrm{g}}(t) \begin{pmatrix} \cos(x_{10}) \\ \cos(x_{10} - \frac{2\pi}{3}) \\ \cos(x_{10} - \frac{4\pi}{3}) \end{pmatrix}^{\top} \begin{bmatrix} \boldsymbol{I}_3 \\ \boldsymbol{J}_{\Sigma} \end{bmatrix} \boldsymbol{x}_{7\text{-}9}}_{=:\boldsymbol{h}_{\mathrm{abc}}(\boldsymbol{x}, t) \in \mathbb{R}^2} = \begin{pmatrix} p_{\mathrm{pcc}}(\boldsymbol{i}_{\mathrm{f}}^{abc}, t) \\ q_{\mathrm{pcc}}(\boldsymbol{i}_{\mathrm{f}}^{abc}, t) \end{pmatrix}
\end{aligned}\right\} \tag{21}$$

with initial values $\boldsymbol{x}(0) = \Big( (\boldsymbol{i}_{\mathrm{s},0}^{abc})^{\top},\ \omega_{\mathrm{m},0},\ \phi_{\mathrm{m},0},\ u_{\mathrm{dc},0},\ (\boldsymbol{i}_{\mathrm{f},0}^{abc})^{\top},\ \phi_{\mathrm{g},0},\ \beta_{\diamondsuit,0} \Big)^{\top}$. Note that, for brevity and clarity, the argument $t$ is only shown for external (purely time-varying) signals like wind $v_{\mathrm{w}}(\cdot)$, carrier signal $c_{\wedge}(\cdot)$ of modulator, grid amplitude $\hat{u}_{\mathrm{g}}(\cdot)$ and grid angular frequency $\omega_{\mathrm{g}}(\cdot)$.

### *C. Electrical system in (simplified) synchronously rotating $(d, q)$-reference frame*

Due to the star-connection of machine (stator) windings and grid-side network, the sums of stator and filter currents are zero for all time, i.e. $i_{\mathrm{s}}^{a}(t) + i_{\mathrm{s}}^{b}(t) + i_{\mathrm{s}}^{c}(t) = i_{\mathrm{f}}^{a}(t) + i_{\mathrm{f}}^{b}(t) + i_{\mathrm{f}}^{c}(t) = 0$ for all $t \geq 0$, respectively. Hence, only two currents on machine and grid side are free to choose. Applying the Clarke-Park transformation (3) (with transformation angle $\phi_{\mathrm{p}} = n_{\mathrm{p}}\phi_{\mathrm{m}}$ and $\phi_{\mathrm{p}} = \phi_{\mathrm{g}}$ on machine and grid side, resp.) to the machine-side and grid-side entries in (21) yields a representation of the respective models in the simplified synchronously rotating $(d, q)$-reference frame. To detect the grid voltage angle $\phi_{\mathrm{g}}$ used in Clarke-Park transformation $\boldsymbol{T}_{\mathrm{cp}}(\phi_{\mathrm{g}})$, a phase-locked loop is usually used [1]. For details see [44] and [48].

[4] In Europe, the grid frequency $f_{\mathrm{g}}$ (hence, $\omega_{\mathrm{g}} = 2\pi f_{\mathrm{g}}$) must remain within the frequency band $50\,\mathrm{Hz} \pm 0.5\,\mathrm{Hz}$ to ensure grid stability (see [59, pp. 13,20,27]).

*1) Machine-side dynamics: Electrical machine (generator) and drive train:* For permanent-magnet flux linkage orientation, i.e. $\phi_{\mathrm{p}}(t) = n_{\mathrm{p}}\phi_{\mathrm{m}}(t)$, the PMSG dynamics simplify. The simplified PMSG dynamics in the $(d,q)$-reference frame are given by [44]

$$\left.\begin{aligned}
\tfrac{\mathrm{d}}{\mathrm{d}t}\boldsymbol{i}_{\mathrm{s}}^{dq}(t) &= L_{\mathrm{s}}^{-1}\Big[\boldsymbol{u}_{\mathrm{s}}^{dq}(t) - R_{\mathrm{s}}\boldsymbol{i}_{\mathrm{s}}^{dq}(t) - n_{\mathrm{p}}\omega_{\mathrm{m}}(t)\boldsymbol{J}\Big(L_{\mathrm{s}}\boldsymbol{i}_{\mathrm{s}}^{dq}(t) + \underbrace{\left(\begin{smallmatrix}\frac{3}{2}\kappa\widehat{\psi}_{\mathrm{pm}}\\ 0\end{smallmatrix}\right)}_{=:\boldsymbol{\psi}_{\mathrm{pm}}^{dq}}\Big)\Big] &&, \boldsymbol{i}_{\mathrm{s}}^{dq}(0) = \boldsymbol{T}_{\mathrm{cp}}(n_{\mathrm{p}}\phi_{\mathrm{m}}(0))\boldsymbol{i}_{\mathrm{s},0}^{abc}\\
\tfrac{\mathrm{d}}{\mathrm{d}t}\omega_{\mathrm{m}}(t) &= \tfrac{1}{\Theta}\Bigg[\frac{m_{\mathrm{t}}\big(v_{\mathrm{w}}(t),\beta(t),\omega_{\mathrm{m}}(t)\big)}{g_{\mathrm{r}}} + \underbrace{n_{\mathrm{p}}\tfrac{2}{3\kappa^2}\boldsymbol{i}_{\mathrm{s}}^{dq}(t)^\top\boldsymbol{J}\boldsymbol{\psi}_{\mathrm{pm}}^{dq}}_{=:m_{\mathrm{m}}\left(\boldsymbol{i}_{\mathrm{s}}^{dq}(t)\right)}\Bigg] &&, \omega_{\mathrm{m}}(0) = \omega_{\mathrm{m},0}\\
\tfrac{\mathrm{d}}{\mathrm{d}t}\phi_{\mathrm{m}}(t) &= \omega_{\mathrm{m}}(t) &&, \phi_{\mathrm{m}}(0) = \phi_{\mathrm{m},0}
\end{aligned}\right\} \quad (22)$$

where $L_{\mathrm{s}} := \frac{3}{2}L_{\mathrm{s,m}} + L_{\mathrm{s},\sigma}$. Note that, for the considered isotropic PMSM, the machine torque is independent of the $d$-current component and simplifies to [48, p. 534]

$$m_{\mathrm{m}}\big(\boldsymbol{i}_{\mathrm{s}}^{dq}\big) = m_{\mathrm{m}}\big(i_{\mathrm{s}}^{q}\big) = n_{\mathrm{p}}\tfrac{2}{3\kappa^2}\tfrac{3}{2}\kappa\,\widehat{\psi}_{\mathrm{pm}}i_{\mathrm{s}}^{q} = \tfrac{n_{\mathrm{p}}}{\kappa}\widehat{\psi}_{\mathrm{pm}}i_{\mathrm{s}}^{q} \quad\Longrightarrow\quad i_{\mathrm{s,ref}}^{q} = \tfrac{\kappa}{n_{\mathrm{p}}\widehat{\psi}_{\mathrm{pm}}}m_{\mathrm{m,ref}} \;\text{ and }\; i_{\mathrm{s,ref}}^{d} = 0, \quad (23)$$

which allows to compute the required reference currents $i_{\mathrm{s,ref}}^{d} = 0$ (to reduce copper losses) and $i_{\mathrm{s,ref}}^{q}$ for given reference torque $m_{\mathrm{m,ref}}$ (in N m).

*2) Power electronics and DC-link dynamics (back-to-back converter):* Note that

$$\forall\,\phi_{\mathrm{p}} \in \mathbb{R}\colon\quad \boldsymbol{T}_{\mathrm{cp}}(\phi_{\mathrm{p}})^{-\top}\boldsymbol{T}_{\mathrm{cp}}(\phi_{\mathrm{p}})^{-1} = \tfrac{2}{3\kappa^2}\boldsymbol{I}_2 \quad\text{and}\quad \boldsymbol{T}_{\mathrm{cp}}(\phi_{\mathrm{p}})\boldsymbol{T}_{\mathrm{vsc}} = \boldsymbol{T}_{\mathrm{cp}}(\phi_{\mathrm{p}}). \quad (24)$$

Hence, the output voltages of the VSCs in the $(d,q)$-reference frame are given by

$$\begin{aligned}
\forall\phi_{\mathrm{p}} \in \{n_{\mathrm{p}}\phi_{\mathrm{m}}, \phi_{\mathrm{g}}\}\ \forall \mathrm{z} \in \{\mathrm{s},\mathrm{f}\}\colon\ \boldsymbol{u}_{\mathrm{z}}^{dq}(\boldsymbol{u}_{\mathrm{z,ref}}^{dq}, u_{\mathrm{dc}}, \phi_{\mathrm{p}}) &:= \boldsymbol{T}_{\mathrm{cp}}(\phi_{\mathrm{p}})\boldsymbol{u}_{\mathrm{z}}^{abc}(\boldsymbol{s}_{\mathrm{z}}^{abc}, u_{\mathrm{dc}}) \overset{(15),(24)}{=} u_{\mathrm{dc}}\boldsymbol{T}_{\mathrm{cp}}(\phi_{\mathrm{p}})\boldsymbol{s}_{\mathrm{z}}^{abc} && (25)\\
&\overset{(17)}{=} u_{\mathrm{dc}}\boldsymbol{T}_{\mathrm{cp}}(\phi_{\mathrm{p}})\boldsymbol{\sigma}\Big(\frac{\boldsymbol{T}_{\mathrm{cp}}(\phi_{\mathrm{p}})^{-1}\boldsymbol{u}_{\mathrm{z,ref}}^{dq}}{u_{\mathrm{dc}}/2} - \boldsymbol{1}_3 c_{\wedge}(t)\Big), && (26)
\end{aligned}$$

which allows to derive the DC-link dynamics as follows

$$\begin{aligned}
\tfrac{\mathrm{d}}{\mathrm{d}t}\,u_{\mathrm{dc}}(t) &\overset{(3),(16)}{=} \tfrac{1}{C_{\mathrm{dc}}}\Big(-\boldsymbol{i}_{\mathrm{s}}^{dq}(t)^\top\boldsymbol{T}_{\mathrm{cp}}\big(n_{\mathrm{p}}\phi_{\mathrm{m}}(t)\big)^{-\top}\boldsymbol{s}_{\mathrm{s}}^{abc}(t) - \boldsymbol{i}_{\mathrm{f}}^{dq}(t)^\top\boldsymbol{T}_{\mathrm{cp}}\big(\phi_{\mathrm{g}}(t)\big)^{-\top}\boldsymbol{s}_{\mathrm{f}}^{abc}(t)\Big),\\
&\overset{(25),(24)}{=} \tfrac{1}{C_{\mathrm{dc}}u_{\mathrm{dc}}}\Big(-\underbrace{\tfrac{2}{3\kappa^2}\boldsymbol{i}_{\mathrm{s}}^{dq}(t)^\top\boldsymbol{u}_{\mathrm{s}}^{dq}(t)}_{=p_{\mathrm{s}}(t)} - \underbrace{\tfrac{2}{3\kappa^2}\boldsymbol{i}_{\mathrm{f}}^{dq}(t)^\top\boldsymbol{u}_{\mathrm{f}}^{dq}(t)}_{=p_{\mathrm{f}}(t)}\Big) && (27)\\
&\overset{(26)}{=} \tfrac{2}{3\kappa^2 C_{\mathrm{dc}}}\Big(-\boldsymbol{i}_{\mathrm{s}}^{dq}(t)^\top\boldsymbol{T}_{\mathrm{cp}}(n_{\mathrm{p}}\phi_{\mathrm{m}})\boldsymbol{\sigma}\Big(\frac{\boldsymbol{T}_{\mathrm{cp}}(n_{\mathrm{p}}\phi_{\mathrm{m}})^{-1}\boldsymbol{u}_{\mathrm{s,ref}}^{dq}}{u_{\mathrm{dc}}/2} - \boldsymbol{1}_3 c_{\wedge}(t)\Big)\\
&\qquad -\boldsymbol{i}_{\mathrm{f}}^{dq}(t)^\top\boldsymbol{T}_{\mathrm{cp}}(\phi_{\mathrm{g}})\boldsymbol{\sigma}\Big(\frac{\boldsymbol{T}_{\mathrm{cp}}(\phi_{\mathrm{g}})^{-1}\boldsymbol{u}_{\mathrm{f,ref}}^{dq}}{u_{\mathrm{dc}}/2} - \boldsymbol{1}_3 c_{\wedge}(t)\Big)\Big). && (28)
\end{aligned}$$

*3) Grid-side dynamics (filter, PCC, and grid):* The grid-side dynamics also simplify. For grid voltage orientation, i.e. $\phi_{\mathrm{p}} = \phi_{\mathrm{g}}$, the grid-side system in the $(d,q)$-reference frame is given by [44]

$$\tfrac{\mathrm{d}}{\mathrm{d}t}\boldsymbol{i}_{\mathrm{f}}^{dq}(t) = \tfrac{1}{L_{\mathrm{f}}}\Big[\boldsymbol{u}_{\mathrm{f}}^{dq}(t) - R_{\mathrm{f}}\boldsymbol{i}_{\mathrm{f}}^{dq}(t) - \omega_{\mathrm{g}}(t)L_{\mathrm{f}}\boldsymbol{J}\boldsymbol{i}_{\mathrm{f}}^{dq}(t) - \underbrace{\left(\begin{smallmatrix}\frac{3}{2}\kappa\,\hat{u}_{\mathrm{g}}(t)\\ 0\end{smallmatrix}\right)}_{=:\boldsymbol{u}_{\mathrm{g}}^{dq}(t)}\Big],\quad \boldsymbol{i}_{\mathrm{f}}^{dq}(0) = \boldsymbol{T}_{\mathrm{cp}}(\phi_{\mathrm{g}}(0))\boldsymbol{i}_{\mathrm{f},0}^{abc} \in \mathbb{R}^2, \quad (29)$$

*4) Power output:* In the $(d,q)$-reference frame, active and reactive instantaneous powers $p_{\mathrm{pcc}}$ (in W) and $q_{\mathrm{pcc}}$ (in var) at the PCC simplify to (cf. [44] and [48, Chap. 14])

$$p_{\mathrm{pcc}}(\boldsymbol{i}_{\mathrm{f}}^{dq}, t) = \tfrac{2}{3\kappa^2}\boldsymbol{u}_{\mathrm{g}}^{dq}(t)^\top\boldsymbol{i}_{\mathrm{f}}^{dq} \overset{(29)}{=} \tfrac{1}{\kappa}\,\hat{u}_{\mathrm{g}}(t)i_{\mathrm{f}}^{d} \quad\text{and}\quad q_{\mathrm{pcc}}(\boldsymbol{i}_{\mathrm{f}}^{dq}, t) = \tfrac{2}{3\kappa^2}\boldsymbol{u}_{\mathrm{g}}^{dq}(t)^\top\boldsymbol{J}\boldsymbol{i}_{\mathrm{f}}^{dq} \overset{(29)}{=} -\tfrac{1}{\kappa}\,\hat{u}_{\mathrm{g}}(t)i_{\mathrm{f}}^{q}. \quad (30)$$

with $\boldsymbol{u}_{\mathrm{g}}^{dq}$ and $\hat{u}_{\mathrm{g}}$ as in (29) and $\kappa \in \{\frac{2}{3}, \sqrt{\frac{2}{3}}\}$.

*5) Overall dynamics in nonlinear state-space representation:* In the $(d,q)$-reference frame, the overall model is of nine-th order. Note that switching is still considered. For (reduced) state vector

$$\boldsymbol{x} := \Big(\underbrace{(x_1,\, x_2)}_{=:\boldsymbol{x}_{1\text{-}2}^\top},\ x_3,\ x_4,\ x_5,\ \underbrace{(x_6,\, x_7)}_{=:\boldsymbol{x}_{6\text{-}7}^\top},\ x_8,\ x_9\Big)^\top := \big((\boldsymbol{i}_{\mathrm{s}}^{dq})^\top,\ \omega_{\mathrm{m}},\ \phi_{\mathrm{m}},\ u_{\mathrm{dc}},\ (\boldsymbol{i}_{\mathrm{f}}^{dq})^\top,\ \phi_{\mathrm{g}},\ \beta_{\diamond}\big)^\top \in \mathbb{R}^9, \quad (31)$$

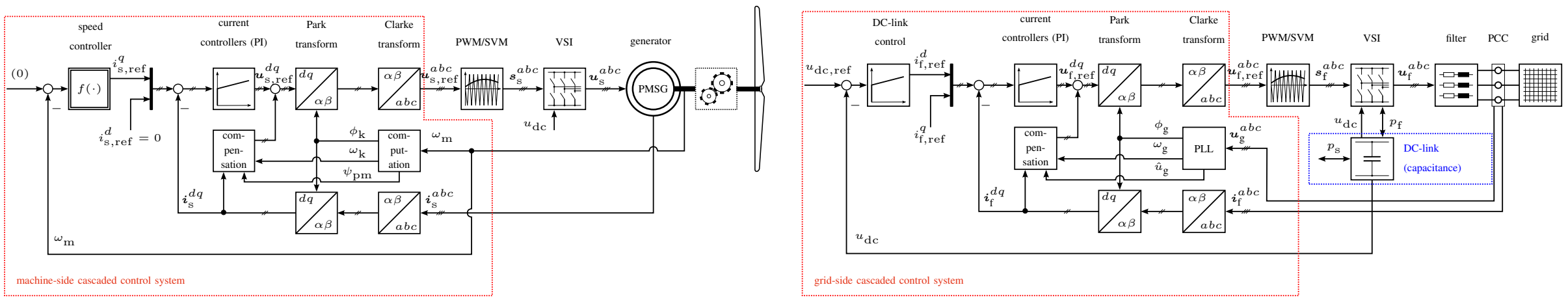

(a) *Machine-side control system (Regime II).* (b) *Grid-side control system.*

Figure 7: *Overview of cascaded wind turbine control systems.*

and (reduced) control input

$$\boldsymbol{u} := \Big(\underbrace{(u_1,\, u_2)}_{=:\boldsymbol{u}_{1\text{-}2}^\top},\ \underbrace{(u_3,\, u_4)}_{=:\boldsymbol{u}_{3\text{-}4}^\top},\ u_5\Big)^\top := \Big((\boldsymbol{u}_{\mathrm{s,ref}}^{dq})^\top,\ (\boldsymbol{u}_{\mathrm{f,ref}}^{dq})^\top, \beta_{\mathrm{ref}}\Big)^\top \in \mathbb{R}^5, \tag{32}$$

the overall (but reduced) nonlinear system dynamics with output are given by the following ninth-order ordinary differential equation

$$\left.\begin{aligned}
\frac{\mathrm{d}}{\mathrm{d}t}\boldsymbol{x} &= \underbrace{\begin{pmatrix}
\frac{1}{L_\mathrm{s}}\Big[x_5 \boldsymbol{T}_{\mathrm{cp}}(n_\mathrm{p}x_4)\boldsymbol{\sigma}\Big(\frac{\boldsymbol{T}_{\mathrm{cp}}(n_\mathrm{p}x_4)^{-1}\boldsymbol{u}_{1\text{-}2}}{x_5/2} - \boldsymbol{1}_3 c_\wedge(t)\Big) - R_\mathrm{s}\boldsymbol{x}_{1\text{-}2} - n_\mathrm{p}x_3\boldsymbol{J}\Big(L_\mathrm{s}\boldsymbol{x}_{1\text{-}2} + \begin{pmatrix}\frac{3}{2}\kappa\widehat{\psi}_{\mathrm{pm}}\\ 0\end{pmatrix}\Big)\Big]\\
\frac{1}{\Theta}\Big[\varrho\,\pi\, r_\mathrm{t}^2\, v_\mathrm{w}(t)^3\,\frac{c_\mathrm{p}\big(r_\mathrm{t}x_3/(g_\mathrm{r}v_\mathrm{w}(t)),\, \mathrm{sat}_{0^\circ}^{90^\circ}(x_9)\big)}{2x_3} + \frac{n_\mathrm{p}}{\kappa}\widehat{\psi}_{\mathrm{pm}}x_2\Big]\\
x_3\\
\frac{2}{3\kappa^2 C_{\mathrm{dc}}}\Big[-\boldsymbol{x}_{1\text{-}2}^\top\boldsymbol{T}_{\mathrm{cp}}(n_\mathrm{p}x_4)\boldsymbol{\sigma}\Big(\frac{\boldsymbol{T}_{\mathrm{cp}}(n_\mathrm{p}x_4)^{-1}\boldsymbol{u}_{1\text{-}2}}{x_5/2} - \boldsymbol{1}_3 c_\wedge(t)\Big) - \boldsymbol{x}_{6\text{-}7}^\top\boldsymbol{T}_{\mathrm{cp}}(x_{10})\boldsymbol{\sigma}\Big(\frac{\boldsymbol{T}_{\mathrm{cp}}(x_{10})^{-1}\boldsymbol{u}_{3\text{-}4}}{x_5/2} - \boldsymbol{1}_3 c_\wedge(t)\Big)\Big]\\
\frac{1}{L_\mathrm{f}}\Big[x_5\boldsymbol{T}_{\mathrm{cp}}(x_{10})\boldsymbol{\sigma}\Big(\frac{\boldsymbol{T}_{\mathrm{cp}}(x_{10})^{-1}\boldsymbol{u}_{3\text{-}4}}{x_5/2} - \boldsymbol{1}_3 c_\wedge(t)\Big) - R_\mathrm{f}\boldsymbol{x}_{6\text{-}7} - \omega_\mathrm{g}(t)L_\mathrm{f}\boldsymbol{J}\boldsymbol{x}_{6\text{-}7} - \begin{pmatrix}\frac{3}{2}\kappa\,\hat{u}_\mathrm{g}(t)\\ 0\end{pmatrix}\Big]\\
\omega_\mathrm{g}(t)\\
\mathrm{sat}_{-\dot{\beta}_{\max}}^{\dot{\beta}_{\max}}\Big(\frac{1}{T_\beta}\big(-\mathrm{sat}_{0^\circ}^{90^\circ}(x_9) + u_5\big)\Big)
\end{pmatrix}}_{=:\boldsymbol{f}_{\mathrm{dq}}(\boldsymbol{x},\boldsymbol{u},t)\in\mathbb{R}^9}\\
\boldsymbol{y} &= \underbrace{\frac{1}{\kappa}\begin{pmatrix}\hat{u}_\mathrm{g}(t)\\ 0\end{pmatrix}^\top\begin{bmatrix}\boldsymbol{I}_2\\ \boldsymbol{J}\end{bmatrix}\boldsymbol{x}_{6\text{-}7}}_{=:\boldsymbol{h}_{\mathrm{dq}}(\boldsymbol{x},t)\in\mathbb{R}^2} = \begin{pmatrix}p_{\mathrm{pcc}}(\boldsymbol{i}_\mathrm{f}^{dq}, t)\\ q_{\mathrm{pcc}}(\boldsymbol{i}_\mathrm{f}^{dq}, t)\end{pmatrix}
\end{aligned}\right\} \tag{33}$$

with initial value $\boldsymbol{x}(0) = \Big((\boldsymbol{T}_{\mathrm{cp}}(n_\mathrm{p}\phi_\mathrm{m}(0))\boldsymbol{i}_{\mathrm{s},0}^{abc})^\top,\ \omega_{\mathrm{m},0},\ \phi_{\mathrm{m},0},\ u_{\mathrm{dc},0},\ (\boldsymbol{T}_{\mathrm{cp}}(\phi_\mathrm{g}(0))\boldsymbol{i}_{\mathrm{f},0}^{abc})^\top,\ \phi_{\mathrm{g},0},\ \beta_{\diamondsuit,0}\Big)^\top$. Again, the argument $t$ is only shown for wind $v_\mathrm{w}(\cdot)$, modulator carrier signal $c_\wedge(\cdot)$, grid voltage amplitude $\hat{u}_\mathrm{g}(\cdot)$ and grid angular frequency $\omega_\mathrm{g}(\cdot)$. Note that, in view of the star connection on machine and grid side, the models (21) and (33) are equivalent.

## IV. Control systems and operation management

### A. Controllers

In this section, the individual controllers of the cascaded control system are described. The overall control system is shown in Fig. 7 for (a) machine side and (b) grid side.

*1) Machine-side and grid-side current controllers ($\mathrm{z} \in \{\mathrm{s}, \mathrm{f}\}$):* The current closed-loop systems on machine/stator ($\mathrm{z} = \mathrm{s}$) and grid/filter ($\mathrm{z} = \mathrm{f}$) side consist of two PI controllers (usually implemented in the $(d, q)$-reference frame), two disturbance compensation feedforward controllers, and the respective current dynamics as presented above. The applied control action consists of two parts, for $\mathrm{z} \in \{\mathrm{s,f}\}$, as follows

$$\boldsymbol{u}_{\mathrm{z,ref}}^{dq}(t) = \underbrace{\boldsymbol{u}_{\mathrm{z,pi}}^{dq}(t)}_{\text{PI controller output}} + \underbrace{\boldsymbol{u}_{\mathrm{z,comp}}^{dq}(t)}_{\text{disturbance compensation}}. \tag{34}$$

Details can be found in e.g. [62, Sec. 7.1.1] or [44] (with similar notation as here). Hence, the voltage reference $\boldsymbol{u}_{\mathrm{z,ref}}^{dq} = (u_{\mathrm{z,ref}}^d,\, u_{\mathrm{z,ref}}^q)^\top$ – the control input to the VSC – is the sum of the disturbance compensation $\boldsymbol{u}_{\mathrm{z,comp}}^{dq} = (u_{\mathrm{z,comp}}^d,\, u_{\mathrm{z,comp}}^q)^\top$ and the output $\boldsymbol{u}_{\mathrm{z,pi}}^{dq} = (u_{\mathrm{z,pi}}^d,\, u_{\mathrm{z,pi}}^q)^\top$ of the PI controller(s). The goal of the disturbance compensation is to obtain (almost) *decoupled* current dynamics for controller design in the $(d,q)$-reference frame. Therefore, depending on the application, the coupling term [44], [63] are, for $\mathrm{z}\in\{\mathrm{s,f}\}$, given by

$$\boldsymbol{u}_{\mathrm{z,dist}}^{dq}(t) := \begin{cases} -n_{\mathrm{p}}\omega_{\mathrm{m}}(t)\boldsymbol{J}\big(L_{\mathrm{s}}\boldsymbol{i}_{\mathrm{s}}^{dq}(t)+\boldsymbol{\psi}_{\mathrm{pm}}^{dq}\big), & \text{for PMSMs as in (22)} \\ -\omega_{\mathrm{g}}(t)L_{\mathrm{f}}\boldsymbol{J}\boldsymbol{i}_{\mathrm{f}}^{dq}(t)-\boldsymbol{u}_{\mathrm{g}}^{dq}(t), & \text{for RL-filter \& grid as in (29)}, \end{cases} \tag{35}$$

which can be (roughly) compensated for by introducing the following feedforward control action $\boldsymbol{u}_{\mathrm{z,comp}}^{dq}=-\boldsymbol{u}_{\mathrm{z,dist}}^{dq}$.

It is well known that PI(D) controllers in presence of input saturation may exhibit integral windup (in particular for large initial errors) leading to large overshoots and/or oscillations in the closed-loop system response (see, e.g., [64], [65]). Due to the limited DC-link voltage $u_{\mathrm{dc}}$ (in V), the output of the VSC is constrained by the saturation level

$$\widehat{u}(u_{\mathrm{dc}}) \in \left[\tfrac{u_{\mathrm{dc}}}{2}, \tfrac{2u_{\mathrm{dc}}}{3}\right] \tag{36}$$

(in V) which depends on the employed modulation strategy (such as pulse-width modulation (PWM) or space-vector modulation (SVM) with or without over-modulation [54, Sec. 8.4]).

Due to the input saturation, a simple but effective anti-windup strategy (similar to *conditional integration*) is implemented which stops integration of the integral control action if the control input (here $\boldsymbol{u}_{\mathrm{z}}^{dq}$ or $\boldsymbol{u}_{\mathrm{z,ref}}^{dq}$) exceeds the admissible range. For this transition function (2) is combined with the two-input two-output PI controller as follows

$$\left.\begin{aligned} \tfrac{\mathrm{d}}{\mathrm{d}t}\boldsymbol{\xi}_{\mathrm{z}}^{dq}(t) &= f_{\widehat{u},\Delta_{\boldsymbol{\xi}_{\mathrm{z}}}}\big(\|\boldsymbol{u}_{\mathrm{z,ref}}^{dq}(t)\|\big)\,\boldsymbol{e}_{\mathrm{z}}^{dq}(t), \qquad \boldsymbol{\xi}_{\mathrm{z}}^{dq}(0)=\boldsymbol{\xi}_{\mathrm{z},0}^{dq}\in\mathbb{R}^2 \\ \boldsymbol{u}_{\mathrm{z,pi}}^{dq}(t) &= k_{\mathrm{z,p}}\boldsymbol{e}_{\mathrm{z}}^{dq}(t)+k_{\mathrm{z,i}}\,\boldsymbol{\xi}_{\mathrm{z}}^{dq}(t), \end{aligned}\right\} \tag{37}$$

where $\mathrm{z}\in\{\mathrm{s,f}\}$, $\boldsymbol{\xi}_{\mathrm{z}}^{dq}=(\xi_{\mathrm{z}}^{d},\,\xi_{\mathrm{z}}^{q})^\top$ is the integrator output vector of the PI controller, $\boldsymbol{\xi}_{\mathrm{z},0}^{dq}$ is its initial value and $\boldsymbol{e}_{\mathrm{z}}^{dq}=(e_{\mathrm{z}}^{d},\,e_{\mathrm{z}}^{q})^\top=\boldsymbol{i}_{\mathrm{z,ref}}^{dq}-\boldsymbol{i}_{\mathrm{z}}^{dq}$ is the current tracking error, $\Delta_{\boldsymbol{\xi}_{\mathrm{z}}}$ is the transition interval for during anti-windup [48, Sect. 14.4]. The parameters $k_{\mathrm{z,p}}$ and $k_{\mathrm{z,i}}$ are the proportional and integral controller gain, respectively. The controller gains can be tuned, e.g., according to the "Magnitude Optimum criterion" (i.e., $k_{\mathrm{z,p}}=L_{\mathrm{z}}/(2T_{\mathrm{avg}})$ and $k_{\mathrm{z,i}}=R_{\mathrm{z}}/(2T_{\mathrm{avg}})$; see [44], [66]) or any other convenient/preferred tuning rule.

*2) Speed controller (Regime II):* For wind speeds below the nominal wind speed, maximum power point tracking (MPPT) is the desired control objective (cf. Fig. 2). The following nonlinear controller, given by

$$m_{\mathrm{m,ref}}(t) = -\,\mathrm{sat}_{0}^{\overline{m}_{\mathrm{m}}}\left[k_{\mathrm{p}}^{\star}\,\omega_{\mathrm{m}}(t)^2\right] \quad \text{where} \quad k_{\mathrm{p}}^{\star} := \tfrac{\varrho\pi r_{\mathrm{t}}^5}{2g_{\mathrm{r}}^3}\tfrac{c_{\mathrm{p}}(\lambda^\star,\beta^\star)}{(\lambda^\star)^3}, \tag{38}$$

achieves MPPT even without wind speed measurement[5]. The controller (38) requires knowledge of the optimal tip speed ratio $\lambda^\star$ and optimal pitch angle $\beta^\star\geq 0$ for the turbine to extract the maximum available wind power. Moreover, its output is saturated by $\overline{m}_{\mathrm{m}}$ (e.g. by the nominal/rated machine torque). In [68], for a constant wind speed $v_{\mathrm{w}}>0$, it has been shown that the speed closed-loop system (neglecting the underlying current closed-loop system and the pitch control system, i.e. $m_{\mathrm{m,ref}}=m_{\mathrm{m}}$ and $\beta=\beta^\star$) is stable and the optimal tip speed ratio $\lambda^\star$ (or the optimal speed $\omega_{\mathrm{m}}^\star=\lambda^\star\frac{v_{\mathrm{w}}}{r_{\mathrm{t}}}$) is reached asymptotically.

*3) Torque controller and pitch reference controller (Regime III):* When the wind speed exceeds the nominal wind speed of the turbine, i.e. $v_{\mathrm{w}}>v_{\mathrm{w,rated}}$, the wind turbine system operates in regime III where the rotor/machine speed is controlled indirectly by the pitch control system and the machine-side torque control system outputs the nominal generator torque (cf. Fig. 2). The pitch control system allows to reduce the turbine torque (independently of the wind). To reduce mechanical stress, a smooth (continuous) transition between regime II and regime III is crucial which can be established by introducing an outer pitch reference controller cascade which adjusts the pitch angle reference $\beta_{\mathrm{ref}}$ appropriately. Here, an output-saturated PI controller with anti-windup is proposed. Its nonlinear dynamics is given by

$$\left.\begin{aligned} \tfrac{\mathrm{d}}{\mathrm{d}t}\xi_{\beta}(t) &= f_{0^\circ,\Delta_{\xi_\beta}}\Big(-k_{\beta,\mathrm{p}}\underbrace{\big(\omega_{\mathrm{m,rated}}-\omega_{\mathrm{m}}(t)\big)}_{=:e_{\omega_{\mathrm{m}}}(t)}-k_{\beta,\mathrm{i}}\xi_{\beta}(t)\Big)\,e_{\omega_{\mathrm{m}}}(t), \quad \xi_{\beta}(0)=\xi_{\beta,0} \\ \beta_{\mathrm{ref}}(t) &= \mathrm{sat}_{0^\circ}^{90^\circ}\big[k_{\beta,\mathrm{p}}e_{\omega_{\mathrm{m}}}(t)+k_{\beta,\mathrm{i}}\xi_{\beta}(t)\big], \end{aligned}\right\} \tag{39}$$

and depend on speed error $e_{\omega}:=\omega_{\mathrm{m,rated}}-\omega_{\mathrm{m}}$, integrator state $\xi_{\beta}$ with anti-windup transition function $f_{0^\circ,\Delta_{\xi_\beta}}(\cdot)$ as in (2) (with $\widehat{a}=0^\circ$ and $\Delta=\Delta_{\xi_\beta}>0$), proportional gain $k_{\beta,\mathrm{p}}>0$ and integrator gain $k_{\beta,\mathrm{i}}>0$. Note that the output $\beta_{\mathrm{ref}}$ of the PI controller is saturated to the interval $[0^\circ,90^\circ]$. Figure 8 shows the block diagram of the nonlinear PI controller implementation.

*4) DC-link voltage controller and reactive power feedforward controller:* On the grid side, active and reactive power can be fed into the grid. The active power is indirectly controlled by the DC-link voltage controller via the $d$-component of the grid-side currents, whereas the reactive power is controlled by a simple feedforward controller via the $q$-component of the filter current. DC-link voltage control is a non-trivial task, since the DC-link system dynamics might exhibit a non-minimum phase behavior. Therefore, some care must be exercised during the controller tuning leading to a rather conservative design (for more details see [44], [69], [70]). In most cases, a PI controller is implemented for DC-link voltage control. Such a PI controller with anti-windup is given by

$$\left.\begin{aligned} \tfrac{\mathrm{d}}{\mathrm{d}t}\xi_{\mathrm{dc}}(t) &= f_{\hat{\imath},\Delta_{\xi_{\mathrm{dc}}}}\big(\|\boldsymbol{i}^{dq}_{\mathrm{f,ref}}(t)\|\big)\,\big(\underbrace{u_{\mathrm{dc,ref}}(t)-u_{\mathrm{dc}}(t)}_{=:e_{\mathrm{dc}}(t)}\big), \qquad \xi_{\mathrm{dc}}(0)=\xi_{\mathrm{dc},0}\in\mathbb{R} \\ i^{d}_{\mathrm{f,ref}}(t) &= k_{\mathrm{dc,p}}e_{\mathrm{dc}}(t)+k_{\mathrm{dc,i}}\,\xi_{\mathrm{dc}}(t), \end{aligned}\right\} \tag{40}$$

where $f_{\hat{\imath},\Delta_{\xi_{\mathrm{dc}}}}(\cdot)$ is as in (2), $\hat{\imath}$ is the maximally admissible filter current amplitude (e.g. nominal current) and $\Delta_{\xi_{\mathrm{dc}}}$ is the transition interval for anti-windup. For any given reference reactive power $q_{\mathrm{pcc,ref}}$ (provided e.g. by the grid operator), the grid-side reference $q$-current $i^{q}_{\mathrm{f,ref}}$ can be obtained by rearranging (30). The reference current

$$i^{q}_{\mathrm{f,ref}}(t) = -\kappa\frac{q_{\mathrm{pcc,ref}}(t)}{\hat{u}_{\mathrm{g}}(t)} \tag{41}$$

can then be used to feedforward control the desired reactive power.

[5] This implies a perfectly working torque control. If this assumption does not hold, this might impact wind turbine efficiency and power production [67].

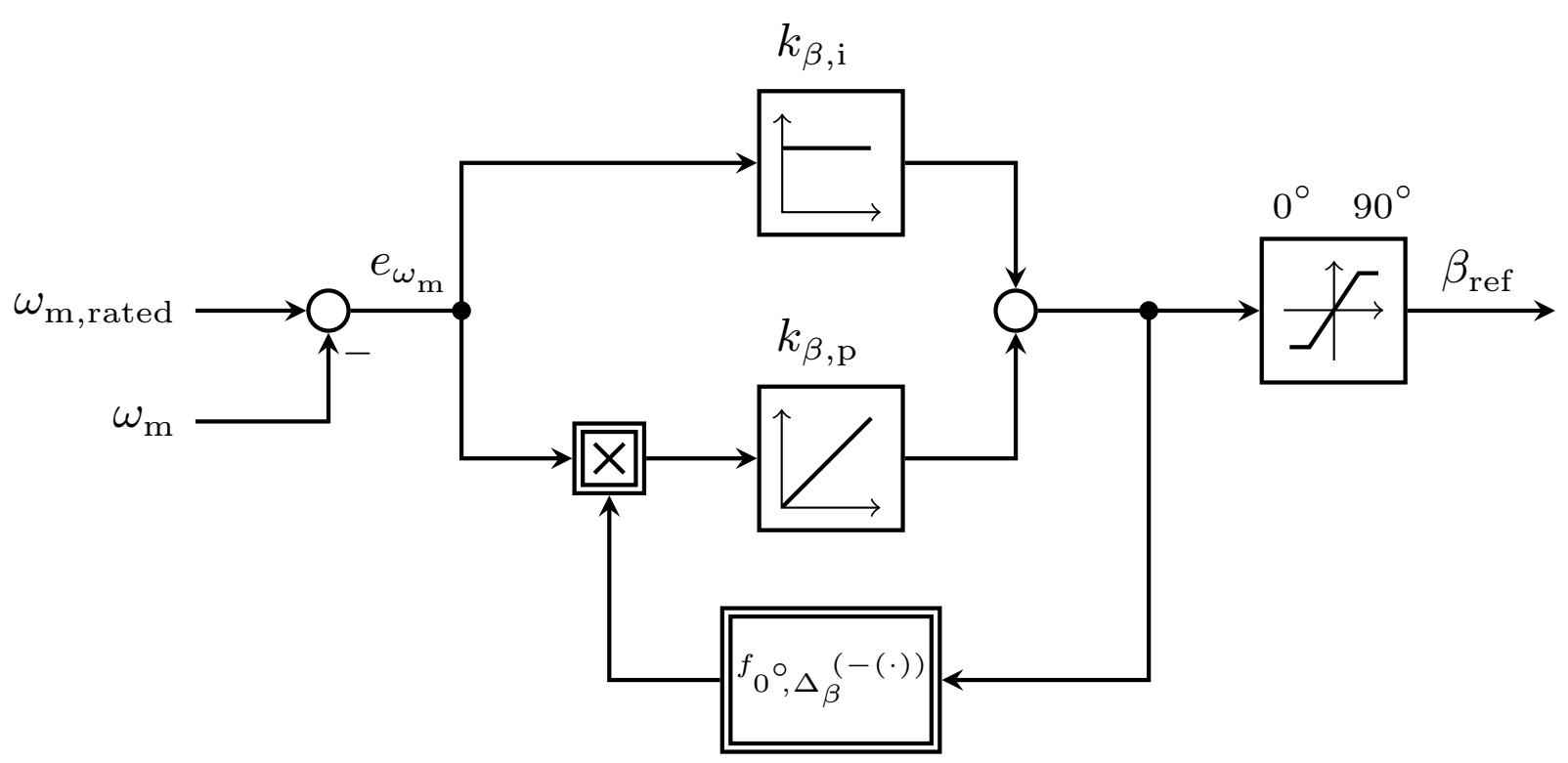

Figure 8: *Nonlinear pitch reference PI controller.*

### B. *Operation management*

The operation management is the high-level control system of the WECS (see Fig. 1). Based on the measured wind speed $\widehat{v}_{\mathrm{w}}$ (rough estimate of the actually incoming wind speed $v_{\mathrm{w}}$), it aligns the rotor perpendicular to the wind direction (yawing; not considered in this paper) and it commands the transitions between the four operation regimes. For example, it triggers startup (transition between regime I $\to$ regime II or regime IV $\to$ regime III) or shutdown (transition between regime II $\to$ regime I or regime III $\to$ regime IV). Moreover, the operation management is the link between single WECSs and the wind park management system or the system operator (TSO). It receives reference values for reactive (in the future, also active) power and propagates them to the underlying control system. Additionally, it also might provide the dc-link voltage reference $u_{\mathrm{dc,ref}}$ for the back-to-back converter or decides wether to perform an emergency shutdown to protect the WECS.

## V. Model Reduction

The presented WECS models in Section III consider all relevant dynamic effects and switching of the power electronic devices. The dynamic models are of eleventh and ninth order for the $(a,b,c)$ and the $(d,q)$-reference frame, respectively. For many studies concerning large-scale power systems or high-level controller design of renewable energy systems for contributing in frequency and voltage stability, these detailed models are not a viable option. Thus, in this section, two models with reduced complexity are developed. In contrast to existing low-complexity WECS models [13]–[15], [27], [28], the model reduction in this paper follows a systematic step by step procedure with well-founded simplification assumptions.

### A. *Non-switching model (nsm)*

In a first simplification step, the explicit switching of the power electronics is neglected. The voltages applied on machine (z = s) and grid side (z = f) by the power electronics can be assumed to be equal to the averaged but saturated output phase voltage vector $\boldsymbol{u}^{abc}_{\mathrm{z}}(t) = \mathbf{sat}_{\widehat{u}}\big(\overline{\boldsymbol{u}}^{abc}_{\mathrm{z}}(t)\big) \approx \mathbf{sat}_{\widehat{u}}\big(\boldsymbol{u}^{abc}_{\mathrm{z,ref}}(t-T_{\mathrm{avg}})\big)$ with $\widehat{u}=\widehat{u}(u_{\mathrm{dc}})$ as in (36) (see also (15) and (18) in Section III-B2 and (1) for a definition of the saturation function $\mathbf{sat}_{\widehat{u}}(\cdot)$). Moreover, in comparison to the other dynamics of the WECS, the time constant $T_{\mathrm{avg}}$ is negligible (as of in the range of microseconds) finally leading to the simplification $\boldsymbol{u}^{abc}_{\mathrm{z}}(t) = \mathbf{sat}_{\widehat{u}}\big(\boldsymbol{u}^{abc}_{\mathrm{z,ref}}(t)\big)$ which can be written as $\boldsymbol{u}^{dq}_{\mathrm{z}}(t) = \mathbf{sat}_{\widehat{u}}\big(\boldsymbol{u}^{dq}_{\mathrm{z,ref}}(t)\big)$ in the $(d,q)$-reference frame (neglecting cross-coupling terms due to the Park transformation [48, p. 521]). Consequently, the power electronics apply instantaneously the requested but possibly saturated voltages. Now, the resultant *non-switching model* can be introduced. Its state and control input vector are

$$\boldsymbol{x} := \Big(\underbrace{(x_1,\, x_2)}_{=:\boldsymbol{x}_{1\text{-}2}^\top},\quad x_3,\quad x_4,\quad x_5,\quad \underbrace{(x_6,\, x_7)}_{=:\boldsymbol{x}_{6\text{-}7}^\top},\quad x_8,\quad x_9\Big)^\top := \big((\boldsymbol{i}_\mathrm{s}^{dq})^\top,\quad \omega_\mathrm{m},\quad u_\mathrm{dc},\quad (\boldsymbol{i}_\mathrm{f}^{dq})^\top,\quad \beta_\diamond\big)^\top \in \mathbb{R}^7, \tag{42}$$

and

$$\boldsymbol{u} := \Big(\underbrace{(u_1,\, u_2)}_{=:\boldsymbol{u}_{1\text{-}2}^\top},\quad \underbrace{(u_3,\, u_4)}_{=:\boldsymbol{u}_{3\text{-}4}^\top},\quad u_5\Big)^\top := \Big((\boldsymbol{u}_\mathrm{s,ref}^{dq})^\top,\quad (\boldsymbol{u}_\mathrm{f,ref}^{dq})^\top, \beta_\mathrm{ref}\Big)^\top \in \mathbb{R}^5, \tag{43}$$

respectively. Its nonlinear dynamics with output are given by

$$\left.\begin{aligned}
\tfrac{\mathrm{d}}{\mathrm{d}t}\boldsymbol{x} &= \underbrace{\begin{pmatrix} L_\mathrm{s}^{-1}\Big[\mathbf{sat}_{\widehat{u}(x_4)}\big(\boldsymbol{u}_{1\text{-}2}\big) - R_\mathrm{s}\boldsymbol{x}_{1\text{-}2}(t) - n_\mathrm{p} x_3 \boldsymbol{J}\Big(L_\mathrm{s}\boldsymbol{x}_{1\text{-}2} + \left(\begin{smallmatrix}\frac{3}{2}\kappa\widehat{\psi}_\mathrm{pm}\\ 0\end{smallmatrix}\right)\Big)\Big] \\ \frac{1}{\Theta}\Big[\varrho\,\pi\, r_\mathrm{t}^2\, v_\mathrm{w}(t)^3 \frac{c_\mathrm{p}\left(r_\mathrm{t}x_3/(g_\mathrm{r}v_\mathrm{w}(t)),\mathrm{sat}_{0^\circ}^{90^\circ}(x_9)\right)}{2x_3} + \frac{n_\mathrm{p}}{\kappa}\widehat{\psi}_\mathrm{pm}x_2\Big] \\ \frac{2}{3\kappa^2\, C_\mathrm{dc}x_4}\Big[-\boldsymbol{x}_{1\text{-}2}^\top\,\mathbf{sat}_{\widehat{u}(x_4)}\big(\boldsymbol{u}_{1\text{-}2}\big) - \boldsymbol{x}_{6\text{-}7}^\top\,\mathbf{sat}_{\widehat{u}(x_4)}\big(\boldsymbol{u}_{3\text{-}4}\big)\Big] \\ \frac{1}{L_\mathrm{f}}\Big[\mathbf{sat}_{\widehat{u}(x_4)}\big(\boldsymbol{u}_{3\text{-}4}\big) - R_\mathrm{f}\boldsymbol{x}_{6\text{-}7} - \omega_\mathrm{g}(t)L_\mathrm{f}\boldsymbol{J}\boldsymbol{x}_{6\text{-}7} - \left(\begin{smallmatrix}\frac{3}{2}\kappa\,\hat{u}_\mathrm{g}(t)\\ 0\end{smallmatrix}\right)\Big] \\ \mathrm{sat}_{-\dot{\beta}_\mathrm{max}}^{\dot{\beta}_\mathrm{max}}\Big(\frac{1}{T_\beta}\big(-\mathrm{sat}_{0^\circ}^{90^\circ}(x_9) + u_5\big)\Big) \end{pmatrix}}_{=:\boldsymbol{f}_\mathrm{nsm}(\boldsymbol{x},\boldsymbol{u},t)\in\mathbb{R}^7} \\
\boldsymbol{y} &= \underbrace{\frac{1}{\kappa}\begin{pmatrix}\hat{u}_\mathrm{g}(t)\\ 0\end{pmatrix}^\top \begin{bmatrix}\boldsymbol{I}_2\\ \boldsymbol{J}\end{bmatrix}\boldsymbol{x}_{6\text{-}7}\cdot}_{=:\boldsymbol{h}_\mathrm{nsm}(\boldsymbol{x},t)\in\mathbb{R}^2} = \begin{pmatrix} p_\mathrm{pcc}\big(\boldsymbol{i}_\mathrm{f}^{dq},t\big) \\ q_\mathrm{pcc}\big(\boldsymbol{i}_\mathrm{f}^{dq},t\big)\end{pmatrix}
\end{aligned}\right\} \tag{44}$$

As before, the argument $t$ is only shown for external signals such as wind $v_\mathrm{w}(\cdot)$ and grid voltage amplitude $\hat{u}_\mathrm{g}(\cdot)$. Note that the (transformation) angles $\phi_\mathrm{m}$ and $\phi_\mathrm{g}$ are not needed for this WECS model. Compared to the detailed model (33) in the $(d,q)$-reference frame, the non-switching model (44) is of seventh order and does not cover switching. As will be shown in Sect. VI, the simulation time and computational complexity of the non-switching model (44) reduces by several orders of magnitude compared to the simulation time of the detailed model (33) (in particular due to the neglected switching).

**Remark V.1.** *The average voltage dynamics could also be considered and are often approximated by a saturated first order-lag system of the following form [48, Chap. 14]*

$$\forall \mathrm{z} \in \{\mathrm{s},\mathrm{f}\}: \qquad \tfrac{\mathrm{d}}{\mathrm{d}t}\boldsymbol{u}_\mathrm{z}^{dq} = \tfrac{1}{T_\mathrm{avg}}\,\boldsymbol{sat}_{\widehat{u}}\big(-\boldsymbol{u}_\mathrm{z}^{dq} + \boldsymbol{u}_\mathrm{z,ref}^{dq}\big), \qquad \boldsymbol{u}_\mathrm{z}^{dq}(0) = \boldsymbol{0}_2.$$

### *B. Reduced-order model (rom)*

In the next simplification step, no switching will be considered and only the dominant dynamics in (44) are identified. Those dynamics are neglected, which do not significantly contribute to the electrical power output of the WECS. In physical systems, the dynamics are related to the intrinsic energy storages in the system [71]. Considering the non-switching model (44) at its nominal (steady state) working point $p_\mathrm{t} = 2\,\mathrm{MW}$ (cf. [44] and Tab. II), the following stored energies can be computed: $\frac{1}{2}\,\Theta\,\omega_\mathrm{m}^2 \approx 23\,960\,\mathrm{kJ}$, $\frac{1}{2}\,C_\mathrm{dc}u_\mathrm{dc}^2 \approx 349.92\,\mathrm{kJ}$, $\frac{1}{2}\frac{2}{3\kappa^2}\,L_\mathrm{s}^q\,(i_\mathrm{s}^q)^2 \approx 2.1\,\mathrm{kJ}$[6], and $\frac{1}{2}\frac{2}{3\kappa^2}\,L_\mathrm{f}\,\|\boldsymbol{i}_\mathrm{f}^{dq}\|^2 \approx 4.39\,\mathrm{kJ}$, which gives the following energy relations

$$\forall\, v_\mathrm{w} \geq v_\mathrm{w,cut-in}: \quad \tfrac{1}{2}\,\Theta\,\omega_\mathrm{m}^2 \gg \tfrac{1}{2}\,C_\mathrm{dc}u_\mathrm{dc}^2 \gg \tfrac{1}{2}\tfrac{2}{3\kappa^2}\,L_\mathrm{f}\|\boldsymbol{i}_\mathrm{f}^{dq}\|^2 \approx \tfrac{1}{2}\tfrac{2}{3\kappa^2}\,L_\mathrm{s}^q\,(i_\mathrm{s}^q)^2. \tag{45}$$

This comparison of the energy contents shows that, the energies stored in the inductances of the generator and filter are rather small compared to the kinetic energy and the DC-link energy. In other words, in the reduced representation, the electrical dynamics of generator and filter will not be modeled explicitly anymore. The electrical dynamics of the currents, their current control loops with underlying voltage source converters are neglected. This implies that (i) the actual currents can be considered to equal their respective reference currents and (ii) the actual torque equals its reference value, i.e.

$$\boldsymbol{i}_\mathrm{s}^{dq} = \boldsymbol{i}_\mathrm{s,ref}^{dq}, \quad \boldsymbol{i}_\mathrm{f}^{dq} = \boldsymbol{i}_\mathrm{f,ref}^{dq} \quad \text{and} \quad m_\mathrm{m} = m_\mathrm{m,ref} \quad \text{in (33) or (44), resp.} \tag{46}$$

Therefore, the reduced-order model has only a three-dimensional state vector

$$\boldsymbol{x} := \big(x_1,\quad x_2,\quad x_3\big)^\top = \big(\omega_\mathrm{m},\quad u_\mathrm{dc},\quad \beta_\diamond\big)^\top \in \mathbb{R}^3 \tag{47}$$

[6]The factor $\frac{1}{2}\frac{2}{3\kappa^2}$ is due to the Clarke transformation factor $\kappa \in \{\frac{2}{3}, \sqrt{\frac{2}{3}}\}$ which scales electrical power and energy by $\frac{2}{3\kappa^2}$ [48, p. 510].

comprising angular velocity $\omega_{\rm m}$, dc-link voltage $u_{\rm dc}$ and pitch angle $\beta_{\diamondsuit}$. The control input vector

$$\boldsymbol{u} := \Big( u_1, \; \underbrace{(u_2, \, u_3)}_{=: \boldsymbol{u}_{2\text{-}3}^\top} \; u_4 \Big)^\top = \Big( m_{\rm m,ref}, \;\; (\boldsymbol{i}_{\rm f,ref}^{dq})^\top, \;\; \beta_{\rm ref} \Big)^\top \in \mathbb{R}^4 \tag{48}$$

of the reduced-order model consists of the reference signals above and the reference pitch angle $\beta_{\rm ref}$. Note that all references come from the speed controller (38), the dc-link voltage controller (40), the reactive power (feedforward) controller (41) and the pitch controller (39) as introduced in Section IV-A.

Although the generator and filter dynamics are negligible, the resistive losses on machine and grid side should still be considered, since those scale with the squared current magnitude, i.e.

$$p_{R_{\rm s}} := \tfrac{2}{3\kappa^2} R_{\rm s} \|\boldsymbol{i}_{\rm s}^{dq}\|^2 = \tfrac{2}{3\kappa^2} R_{\rm s} \|\boldsymbol{i}_{\rm s,ref}^{dq}\|^2 \overset{(23)}{=} \tfrac{2 R_{\rm s}}{3 n_{\rm p}^2 \widehat{\psi}_{\rm pm}^2} m_{\rm m,ref}^2 \;\; \text{and} \;\; p_{R_{\rm f}} := \tfrac{2}{3\kappa^2} R_{\rm f} \|\boldsymbol{i}_{\rm f}^{dq}\|^2 = \tfrac{2}{3\kappa^2} R_{\rm f} \|\boldsymbol{i}_{\rm f,ref}^{dq}\|^2, \tag{49}$$

respectively. Note that the stator losses can be expressed by a nonlinear function of the reference torque instead of the stator currents. The copper losses in (49) will be included into the dc-link dynamics as follows. The dc-link dynamics (16) can be approximated as follows

$$\begin{aligned}
\tfrac{\rm d}{{\rm d}t} u_{\rm dc} &\overset{(44)}{=} \frac{2}{3\kappa^2 C_{\rm dc} u_{\rm dc}} \Big[ -(\boldsymbol{i}_{\rm s}^{dq})^\top \mathbf{sat}_{\widehat{u}}(\boldsymbol{u}_{\rm s,ref}^{dq}) - (\boldsymbol{i}_{\rm f}^{dq})^\top \mathbf{sat}_{\widehat{u}}(\boldsymbol{u}_{\rm f,ref}^{dq}) \Big] \\
&\overset{(27)}{\approx} \frac{2}{3\kappa^2 C_{\rm dc} u_{\rm dc}} \Big[ -(\boldsymbol{i}_{\rm s}^{dq})^\top \boldsymbol{u}_{\rm s}^{dq} - (\boldsymbol{i}_{\rm f}^{dq})^\top \boldsymbol{u}_{\rm f}^{dq} \Big] \\
&\overset{(22),(29)}{=} \frac{2}{3\kappa^2 C_{\rm dc} u_{\rm dc}} \Big[ -(\boldsymbol{i}_{\rm s}^{dq})^\top \big( R_{\rm s} \boldsymbol{i}_{\rm s}^{dq} + n_{\rm p} \omega_{\rm m} \boldsymbol{J} \boldsymbol{\psi}_{\rm pm}^{dq} \big) - (\boldsymbol{i}_{\rm f}^{dq})^\top \Big( R_{\rm f} \boldsymbol{i}_{\rm f}^{dq} + \Big(\begin{smallmatrix} \frac{3}{2}\kappa\, \hat{u}_{\rm g}(t) \\ 0 \end{smallmatrix}\Big) \Big) \Big] && (50) \\
&= \frac{2}{3\kappa^2 C_{\rm dc} u_{\rm dc}} \Big[ -\big( R_{\rm s} \|\boldsymbol{i}_{\rm s}^{dq}\|^2 + \underbrace{n_{\rm p} \omega_{\rm m} (\boldsymbol{i}_{\rm s}^{dq})^\top \boldsymbol{J} \boldsymbol{\psi}_{\rm pm}^{dq}}_{\overset{(22)}{=} \frac{3\kappa^2}{2} \omega_{\rm m} m_{\rm m}} \big) - \big( R_{\rm f} \|\boldsymbol{i}_{\rm f}^{dq}\|^2 + \tfrac{3}{2}\kappa\, \hat{u}_{\rm g}(t) i_{\rm f}^{d} \big) \Big], && (51)
\end{aligned}$$

where, in the next-to-last step, (22) and (29) were solved for $\boldsymbol{u}_{\rm s}^{dq}$ and $\boldsymbol{u}_{\rm f}^{dq}$, respectively and the results were inserted in (50) while $(\boldsymbol{i}_{\rm s}^{dq})^\top \boldsymbol{J} \boldsymbol{i}_{\rm s}^{dq} = 0$, $(\boldsymbol{i}_{\rm f}^{dq})^\top \boldsymbol{J} \boldsymbol{i}_{\rm f}^{dq} = 0$ and the derivative terms with $\frac{\rm d}{{\rm d}t} \boldsymbol{i}_{\rm s}^{dq}$ and $\frac{\rm d}{{\rm d}t} \boldsymbol{i}_{\rm f}^{dq}$ were neglected (recall motivation above).

Finally, by recalling state vector (47) and control input vector (48), and invoking (46), (51) and (49), the nonlinear state-space representation of the reduced-order model with output is obtained. It is given by

$$\left.\begin{aligned}
\tfrac{\rm d}{{\rm d}t} \boldsymbol{x} &= \underbrace{\begin{pmatrix} \frac{1}{\Theta} \Big[ \varrho \, \pi \, r_{\rm t}^2 \, v_{\rm w}(t)^3 \frac{c_{\rm p}\big(r_{\rm t} x_1 / (g_{\rm r} v_{\rm w}(t)), {\rm sat}_{0^\circ}^{90^\circ}(x_3)\big)}{2 x_1} + u_1 \Big] \\ \frac{1}{C_{\rm dc} x_2} \Big[ - \underbrace{x_1 \, u_1}_{\overset{(12)}{=} p_{\rm m} = p_{\rm t} < 0} - \frac{2 R_{\rm s}}{3 n_{\rm p}^2 \widehat{\psi}_{\rm pm}^2} u_1^2 - \underbrace{\tfrac{1}{\kappa} \hat{u}_{\rm g}(t) \, u_2}_{\overset{(30)}{=} p_{\rm pcc}(i_{\rm f}^d, t)} - \tfrac{2}{3\kappa^2} R_{\rm f} \, \|\boldsymbol{u}_{2\text{-}3}\|^2 \Big] \\ {\rm sat}_{-\beta_{\max}}^{\dot{\beta}_{\max}} \Big( \frac{1}{T_\beta} \big( -{\rm sat}_{0^\circ}^{90^\circ}(x_3) + u_4 \big) \Big) \end{pmatrix}}_{=: \boldsymbol{f}_{\rm rom}(\boldsymbol{x}, \boldsymbol{u}, t) \in \mathbb{R}^3} \\
\boldsymbol{y} &= \underbrace{\tfrac{1}{\kappa} \begin{pmatrix} \hat{u}_{\rm g}(t) \\ 0 \end{pmatrix}^\top \begin{bmatrix} \boldsymbol{I}_2 \\ \boldsymbol{J} \end{bmatrix} \boldsymbol{u}_{2\text{-}3}}_{=: \boldsymbol{h}_{\rm rom}(\boldsymbol{u}, t) \in \mathbb{R}^2} \overset{(30),(46)}{=} \begin{pmatrix} p_{\rm pcc}(i_{\rm f,ref}^d, t) \\ q_{\rm pcc}(i_{\rm f,ref}^q, t) \end{pmatrix}
\end{aligned}\right\} \tag{52}$$

with initial values $\boldsymbol{x}(0) = \big( \omega_{\rm m,0}, \;\; u_{\rm dc,0}, \;\; \beta_{\diamondsuit,0} \big)^\top$. Note that via the dc-link dynamics the mechanical power $p_{\rm m} = p_{\rm t} = \omega_{\rm m} m_{\rm m,ref} < 0$ (minus the copper losses) is transfered to the PCC, where the active power $p_{\rm pcc}$ is induced. Again, only wind speed $v_{\rm w}(\cdot)$ and grid voltage amplitude $\hat{u}_{\rm g}(\cdot)$ are considered as external, time-varying signals. The reduced-order model (52) is of third order. Hence, compared to (44), six state variables are additionally eliminated, which reduces the computation and simulation time further (see Sect. VI). For controller design of the reduced-order model, the controllers (38), (40), (41) and (39) presented in Sect. IV-A can still serve as benchmark design.

## VI. Implementation and comparative simulation results

In this section, the full-order model (33) from Sect. III-C5, the non-switching model (44) from Sect. V-A and the reduced-order model (52) from Sect. V-B were implemented in Matlab/Simulink and Modelica, respectively (for details see Tab. I). Note that, due to the equivalence of the full-order models (21) and (33), only the model in the synchronously rotating reference frame is implemented.

The comparative simulation results of all three implementations are shown in Fig. 9–11. The results will be discussed in the following. The implementations were performed exactly based on the introduced closed-form representations as described in the respective sections. The implementation parameters are listed in Tab. II. The wind data used was measured at the FINO1 research platform ($54^\circ\,00'\,53,5''\,\mathrm{N}$, $06^\circ\,35'\,15,5''\,\mathrm{E}$) on the 23rd of September 2009 between 8:10 - 08:20 am (with a time resolution of 10 Hz). In all Fig. 9–11, the wind speed profile is shown in the upper most subplot. It is varying around the nominal wind speed (under- and overshooting) leading to an operation of the WECS in regime II *and* regime III. In order to achieve a fair comparison, all models are implemented with identical controllers/tuning for pitch angle, DC-link voltage and angular velocity (see Section IV and Tab. II). The closed-loop systems were fed by the identical external signals such as wind speed $v_\mathrm{w}$ and reactive power reference $q_\mathrm{pcc,ref}$. Since the reduced-order model (52) does not consider current dynamics, no current controllers were implemented.

The comparative simulation results of all implementations are split into three plots:

- Fig. 9 compares dc-link voltage $u_\mathrm{dc}$, mechanical angular velocity $\omega_\mathrm{m}$ and pitch angle $\beta$ of all three models;
- Fig. 10 compares turbine $p_\mathrm{t}$, active $p_\mathrm{pcc}$ and reactive $q_\mathrm{pcc}$ power (at the PCC) and the produced energy $E$ of all three models; and
- Fig. 11 compares machine torque $m_\mathrm{m}$ of all three models and currents $i_\mathrm{s}^q$, $i_\mathrm{f}^d$, & $i_\mathrm{f}^q$ of full-state model and non-switching model.

Quantities of non-switching and reduced-order model are indicated by the additional subscript "nsm" and "rom", respectively. The quantities of the full-state model come without additional subscript.

The upper subplot in Fig. 9 shows the wind speed $v_\mathrm{w}$, its mean value $\bar{v}_\mathrm{w}$ and the nominal wind speed $v_\mathrm{w,rated}$ of the WTS. In the major part of the simulation (mainly around $t = 150...530$s), the wind speed is significantly higher than the nominal wind speed of the turbine. Thus, the averaged wind speed is slightly above the nominal wind speed. In the second subplot, the dc-link voltages are depicted: $u_\mathrm{dc,nsm}$ (of non-switching model) and $u_\mathrm{dc,rom}$ (of reduced-order model) coincide and track their reference value of $5.4\,\mathrm{kV}$ nicely. The voltage $u_\mathrm{dc}$ (of the full-state model) also tracks its reference value but due to the modeled switching behavior of the power converters it is not as smooth as the other dc-link voltages, where the switching is not considered in the models. The third subplot shows the angular velocity of the shaft. The velocities of all three models are almost identical and do only exceed the nominal mechanical velocity $\omega_\mathrm{m,rated}$ for short periods. Hence, pitch and speed control system are working properly. The pitch angles of all three implementations are shown in the last subplot. The three pitch controllers work – as expected – very similar and are only active when the wind speed exceeds its nominal value.

The first subplot of Fig. 10 shows again the wind speed. The second subplot depicts nominal turbine power $p_\mathrm{t,rated}$, turbine power $p_\mathrm{t}$ and active power $p_\mathrm{pcc}$ at the PCC of all models. Within the time interval $[150\,\mathrm{s}, 530\,\mathrm{s})$, where the wind speed is (almost always) higher than the nominal wind speed of the turbine, the WECS is operated in regime III and nominal power is fed into the grid. Within the intervals $[0\,\mathrm{s}, 150\,\mathrm{s})$ and $[530\,\mathrm{s}, 600\,\mathrm{s}]$, the power output undergoes strong fluctuations. The WECS is operated in regime II and follows the rapid variations of the wind power $p_\mathrm{w} \propto v_\mathrm{w}^3$. The principle behavior of all models is similar for both operating regimes. However, the switching behavior becomes only obvious in the active power of the full-state model. The active powers of the non-switching and reduced-order models represent the mean (averaged) output power of the full-order model. The reactive powers at the PCC are shown in the third subplot. The reference reactive power is followed almost instantaneously by all models (independently of being capacitive or inductive reactive power), which underpins the capability of WECSs to contribute to voltage stability of the grid. The switching behavior is again only visible for the full-state model. The fourth subplot illustrates the produced energies of turbine $E_\mathrm{t} := \int p_\mathrm{t}\,\mathrm{d}t$ and induced energy $E_\mathrm{pcc} := \int p_\mathrm{pcc}\,\mathrm{d}t$ at the PCC (as integrals of powers over time). Due to the (copper) losses in generator and filter, the turbine energy $E_\mathrm{t}$ is slightly higher than the electrical energies at the PCC of the three models. The trajectories of $E_\mathrm{pcc}$, $E_\mathrm{pcc,nsm}$ and $E_\mathrm{pcc,rom}$ are almost not distinguishable. Hence, the produced energies match as well.

In the first subplot of Fig. 11, again the wind speed is plotted. The second subplot shows the $q$-component of the stator current $i_\mathrm{s}^q$ for the full-state model and the non-switching model. The reduced-order model does not consider current dynamics. The non-switching model current represents the mean (average) current of the full-state model. In the third subplot, the generator torque is depicted. Its dynamics are similar to those of stator current or (negative) filter current (proportional to active power at the PCC). In regime III, the generator torque is almost constant at its nominal value (see interval $[150\,\mathrm{s}, 530\,\mathrm{s})$). The switching can only be observed in the torque of the full-order model, since it is proportional to the stator current which is directly affected by the switching in the converter. The torques of non-switching and reduced-order model are not distinguishable. The fourth and fifth subplots show the $d$- and $q$-components of the filter currents which are proportional to active power and reactive power at the PCC, respectively. In both subplots, the non-switching models gives again the averaged currents of full-order model. As the currents dynamics are not simulated for the reduced-order model, these currents can not be shown.

Table I compares the simulation environments and the duration of all three implementations. All simulations were performed on the same operating system, only the versions differ. The full-state model was implemented in Matlab/Simulink R2013b 64-bit, and the non-switching and reduced-order model in OpenModelica v1.11.0. All simulations were run with the same fixed-step solver (Runge-Kutta (ode4)). The computational power available for the Matlab/Simulink implementation of the full-state model is significantly higher than the computational power used for the simulation of the non-switching and reduced-order model. The computer with the Intel Xeon CPUs has three times more cores than the computer with the Intel i7 CPUs. Moreover, the frequency of each core is higher. The RAM memory available for the the full-state model is more than 15 times larger than the RAM memory for the other models. However, this does not necessarily have direct impact on the simulation duration. The simulation of the full-state model needs almost 60 times longer (with 2:58 h) than the non-switching model (with 00:03 h) in order to compute active and reactive power over the scenario duration of 00:10 h = 600 s (see e.g. Fig. 9). This is due to the higher complexity of the full-state model and due to the smaller simulation step size ($4 \cdot 10^{-6}$ s) required to simulate the switching behavior of the converters correctly. Neglecting the switching behavior allows the use of larger simulation step sizes ($2 \cdot 10^{-4}$ s). So, the simulation durations of the non-switching and reduced-order model are smaller than 3 min (170 s) and 2 min (119 s), respectively. The simulations of these two models without switching of the power electronics could even be further accelerated by using variable-step solvers.

Table I: *Simulation environments and durations.*

| | full-state model (33) | non-switching model (44) | reduced-order model (52) |
|---|---|---|---|
| Operating system | Windows 10 (Education 64-bit) | Windows 10 (Home 64-bit) | Windows 10 (Home 64-bit) |
| CPU | Intel Xenon E5-1650 v3 (3.50 GHz, 12 CPUs) | Intel Core i7-3630QM (2.40 GHz, 4 CPUs) | Intel Core i7-3630QM (2.40 GHz, 4 CPUs) |
| RAM | 131 072 MB | 8 000 MB | 8 000 MB |
| Simulation software | Matlab Simulink (R2013b 64-bit) | OpenModelica (v1.11.0 64-bit) | OpenModelica (v1.11.0 64-bit) |
| Simulation step size | $4 \cdot 10^{-6}$ s | $2 \cdot 10^{-4}$ s | $2 \cdot 10^{-4}$ s |
| Solver (fixed-step) | Runge-Kutta (ode4) | Runge-Kutta (ode4) | Runge-Kutta (ode4) |
| Simulation duration | 2:58 h (10 680 s) | 0:03 h (170 s) | 0:02 h (119 s) |

## VII. Conclusions

Existing control-oriented models of wind energy conversion systems (WECS) are subject to unstructured simplifications. In this technical note, physical 11th-order and 9th-order state-space models for variable-speed variable-pitch WECSs and a corresponding control and operation management system are presented. The models consider all relevant dynamics which significantly affect the active and reactive power output of the system. Based on the 9th-order model, a 7th-order non-switching model and a reduced 3rd-order model are derived based on a simple reduction procedure. The simplified models are associated with a significant reduction of the computation/simulation time and control design complexity. Extensive simulation studies illustrate that all three models produce plausible and comparable results. Due to the modeled switching in the power converters, the 11th-order and 9th-order physical models exhibit high frequency oscillations in the instantaneous power output. The non-switching 7th-order model and the reduced third-order model provide a time-averaged instantaneous power output which correctly reflects the (average) energy produced by the WECS. Regarding the power output, the conducted simulations suggest no significant difference between the non-switching 7th-order model and the third-order model. Thus, for power flow studies, the third-order model can be assumed to produce sufficiently exact results.

## References

[1] R. Teodorescu, M. Liserre, and P. Rodríguez, *Grid Converters for Photovoltaic and Wind Power Systems*. Chichester, United Kingdom: John Wiley & Sons, Ltd., 2011.

[2] Global Wind Energy Council, "Global wind report 2016," tech. rep., 2016.

[3] P. M. Anderson and A. Bose, "Stability simulation of wind turbine systems," *IEEE Transactions on Power Apparatus and Systems*, no. 12, pp. 3791–3795, 1983.

[4] T. Petru and T. Thiringer, "Modeling of wind turbines for power system studies," *IEEE Transactions on Power Systems*, vol. 17, no. 4, pp. 1132–1139, 2002.

[5] J. G. Slootweg, S. W. H. De Haan, H. Polinder, and W. L. Kling, "General model for representing variable speed wind turbines in power system dynamics simulations," *IEEE Transactions on Power Systems*, vol. 18, no. 1, pp. 144–151, 2003.

[6] J. B. Ekanayake, L. Holdsworth, X. Wu, and N. Jenkins, "Dynamic modeling of doubly fed induction generator wind turbines," *IEEE Transactions on Power Systems*, vol. 18, no. 2, pp. 803–809, 2003.

[7] Y. Lei, A. Mullane, G. Lightbody, and R. Yacamini, "Modeling of the wind turbine with a doubly fed induction generator for grid integration studies," *IEEE Transactions on Energy Conversion*, vol. 21, no. 1, pp. 257–264, 2006.

[8] M. Yin, G. Li, M. Zhou, and C. Zhao, "Modeling of the wind turbine with a permanent magnet synchronous generator for integration," in *Power Engineering Society General Meeting*, pp. 1–6, IEEE, 2007.

[9] M. Behnke, A. Ellis, Y. Kazachkov, T. McCoy, E. Muljadi, W. Price, and J. Sanchez-Gasca, "Development and validation of wecc variable speed wind turbine dynamic models for grid integration studies," tech. rep., National Renewable Energy Laboratory (NREL), U.S. Department of Energy, 2007.

[10] C. Subramanian, D. Casadei, A. Tani, and C. Rossi, "Modeling and simulation of grid connected wind energy conversion system based on a doubly fed induction generator (dfig)," *International Journal of Electrical Energy*, vol. 2, no. 2, 2014.

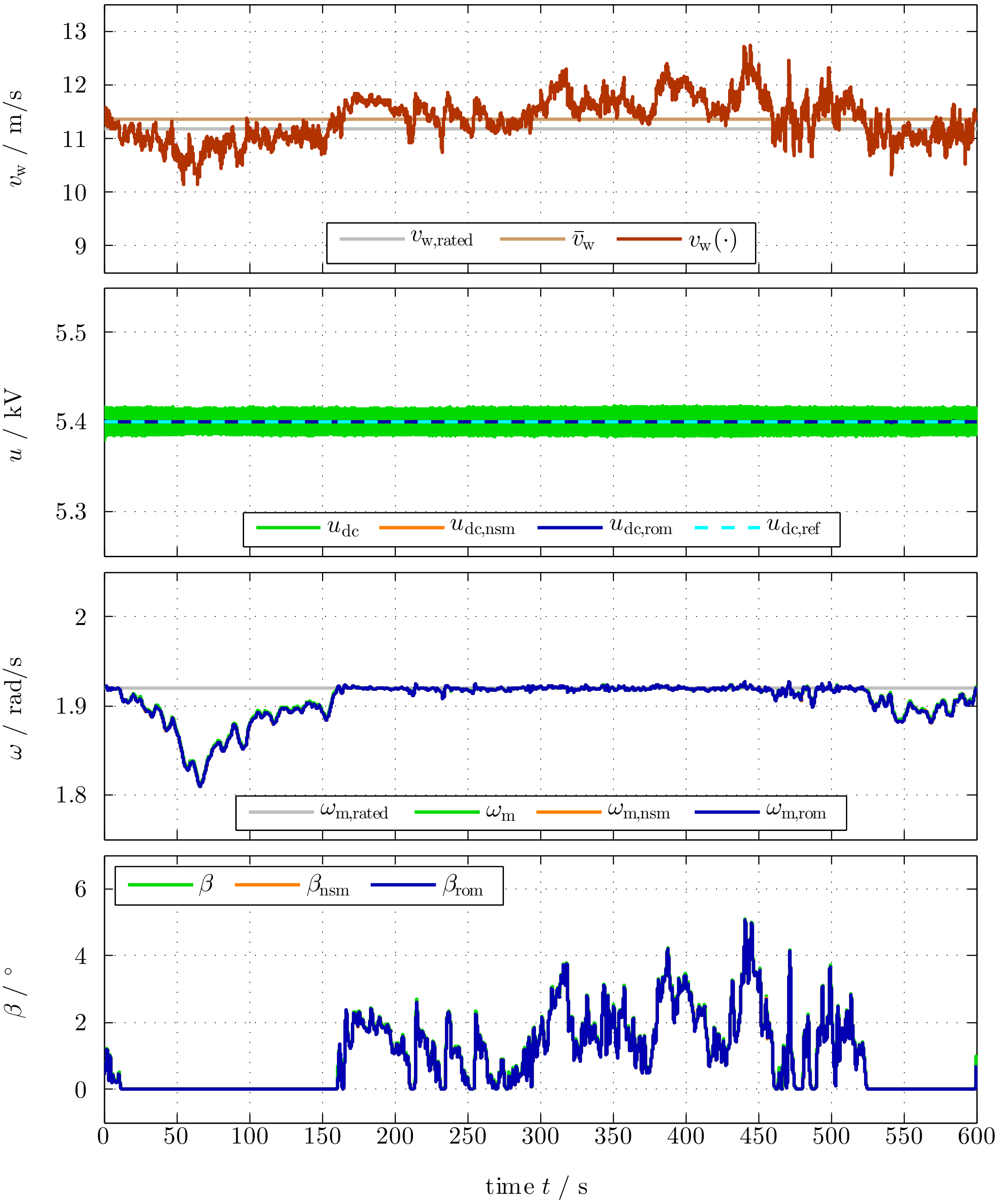

Figure 9: *Comparison of dc-link voltages, mechanical angular velocities, and pitch angles.*

[11] A. Tobías-González, R. Pena-Gallardo, J. Morales-Saldana, and G. Gutiérrez-Urueta, "Modeling of a wind turbine with a permanent magnet synchronous generator for real time simulations," in *2015 IEEE International Autumn Meeting on Power, Electronics and Computing*, pp. 1–6, IEEE, 2015.

[12] F. Huerta, R. L. Tello, and M. Prodanovic, "Real-time power-hardware-in-the-loop implementation of variable-speed wind turbines," *IEEE Transactions on Industrial Electronics*, vol. 64, no. 3, pp. 1893–1904, 2017.

[13] R. Rocha, L. S. M. Martins Filho, and M. V. Bortolus, "Optimal multivariable control for wind energy conversion system – a comparison between $h_2$ and $h_\infty$ controllers," in *Conference on Decision and Control and the European Control Conference*, pp. 7906–7911, IEEE, 2005.

[14] F. Lescher, J. Y. Zhao, and P. Borne, "Switching lpv controllers for a variable speed pitch regulated wind turbine," in *IMACS Multiconference on Computational Engineering in Systems Applications*, pp. 1334–1340, IEEE, 2006.

[15] I. Munteanu, A. I. Bratcu, N.-A. Cutululis, and E. Ceanga, *Optimal Control of Wind Energy Systems*. Springer, 2008.

[16] M. Soliman, O. P. Malik, and D. T. Westwick, "Multiple model predictive control for wind turbines with doubly fed induction generators," *IEEE Transactions on Sustainable Energy*, vol. 2, no. 3, pp. 215–225, 2011.

[17] L. C. Henriksen, N. K. Poulsen, and M. H. Hansen, "Nonlinear model predictive control of a simplified wind turbine," in *IFAC World Congress*, pp. 551–556, Elsevier, 2011.

[18] S. Bououden, M. Chadli, S. Filali, and A. El Hajjaji, "Fuzzy model based multivariable predictive control of a variable speed wind turbine: Lmi approach," *Renewable Energy*, vol. 37, no. 1, pp. 434–439, 2012.

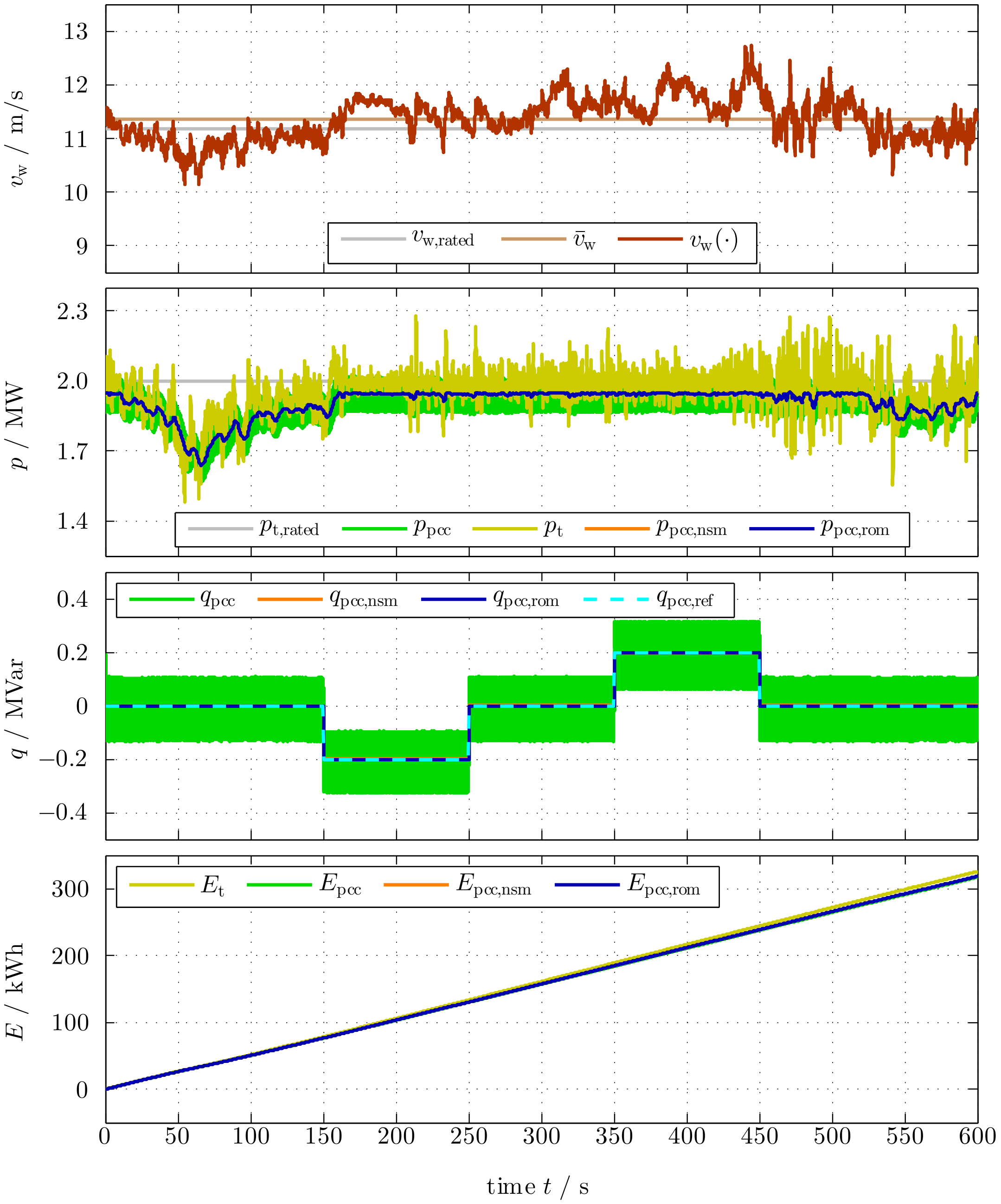

Figure 10: *Comparison of active and reactive powers and produced energy.*

[19] O. Elbeji, M. B. Hamed, and L. Sbita, "Pmsg wind energy conversion system: Modeling and control," *International Journal of Modern Nonlinear Theory and Application*, vol. 3, pp. 88–97, 2014.

[20] S. Yu, K. Emami, T. Fernando, H. H. C. Iu, and K. P. Wong, "State estimation of doubly fed induction generator wind turbine in complex power systems," *IEEE Transactions on Power Systems*, vol. 31, no. 6, pp. 4935–4944, 2016.

[21] M. Zribi, M. Alrifai, and M. Rayan, "Sliding mode control of a variable-speed wind energy conversion system using a squirrel cage induction generator," *Energies*, vol. 10, no. 5, 2017.

[22] G. L. Park, "Planning manual for utility application of wecs," tech. rep., Michigan State University, East Lansing (USA). Division of Engineering Research, 1979.

[23] S. Øye, "Unsteady wake effects caused by pitch-angle changes," in *IEA R&D WECS Joint Action on Aerodynamics of Wind Turbines*, pp. 58–79, 1986.

[24] D. Rekioua, "Wind energy conversion and power electronics modeling," in *Wind Power Electric System: Modeling, Simulation and Control*, pp. 51–76, Springer, 2014.

[25] B. Wu, Y. Lang, N. Zargari, and S. Kouro, *Power Conversion and Control of Wind Energy Systems*. John Wiley & Sons, 2011.

[26] M. Singh and S. Santoso, "Dynamic models for wind turbines and wind power plants," tech. rep., National Renewable Energy Laboratory (NREL), U.S. Department of Energy, 2011.

[27] F. D. Bianchi, R. J. Mantz, and H. De Battista, "Modelling of variable-speed variable-pitch wind energy conversion systems," in *Wind Turbine Control*

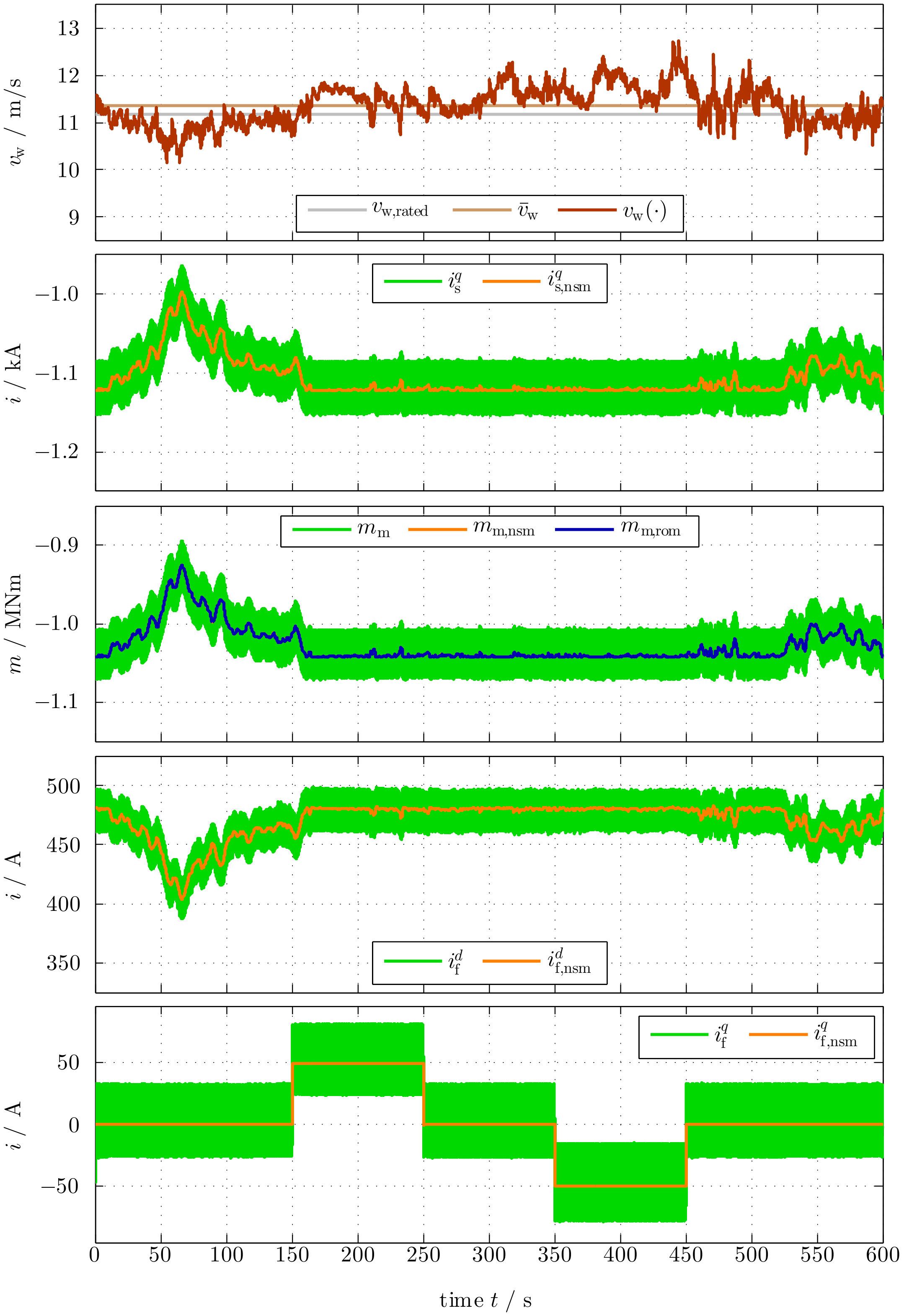

Figure 11: *Comparison of machine torques, and machine-side and filter-side currents (if available in model).*

*Systems: Principles, Modelling and Gain Scheduling Design*, pp. 29–48, Springer, 2007.

[28] J. A. Sanchez, C. Veganzones, S. Martinez, F. Blazquez, N. Herrero, and J. R. Wilhelmi, "Dynamic model of wind energy conversion systems with variable speed synchronous generator and full-size power converter for large-scale power system stability studies," *Renewable Energy*, vol. 33, no. 6, pp. 1186–1198, 2008.

[29] J. G. Slootweg, H. Polinder, and W. L. Kling, "Dynamic modelling of a wind turbine with doubly fed induction generator," in *Power Engineering Society Summer Meeting*, pp. 644–649, IEEE, 2001.

[30] B. H. Chowdhury and S. Chellapilla, "Double-fed induction generator control for variable speed wind power generation," *Electric Power Systems Research*, vol. 76, no. 9, pp. 786–800, 2006.

[31] Z. Sun, H. Wang, and Y. Li, "Modelling and simulation of doubly-fed induction wind power system based on matlab/simulink," in *International Conference on Sustainable Power Generation and Supply*, IET, 2012.

[32] T. Ackermann, *Wind Power in Power Systems*. John Wiley & Sons, 2 ed., 2012.

[33] A. Rolan, A. Luna, G. Vazquez, D. Aguilar, and G. Azevedo, "Modeling of a variable speed wind turbine with a permanent magnet synchronous generator," in *IEEE International Symposium on Industrial Electronics*, pp. 734–739, IEEE, 2009.

[34] M. Singh and S. Santoso, "Dynamic model for full-converter wind turbines employing permanent magnet alternators," in *IEEE Power and Energy Society General Meeting*, IEEE, 2011.

[35] X. Liu and X. Kong, "Nonlinear model predictive control for dfig-based wind power generation," *IEEE Transactions on Automation Science and Engineering*, vol. 11, no. 4, pp. 1046–1055, 2014.

[36] K. O. Merz, "A linear state-space model of an offshore wind turbine, implemented in the stas wind power plant analysis program," tech. rep., SINTEF, 2015.

[37] E. Bolte and M. Landwehr, "Mathematical model of small wind turbines," in *International Conference on Ecological Vehicles and Renewable Energies (EVER)*, pp. 1–6, IEEE, 2014.

[38] S. A. Papathanassiou and M. P. Papadopoulos, "Dynamic behavior of variable speed wind turbines under stochastic wind," *IEEE Transactions on Energy Conversion*, vol. 14, no. 4, pp. 1617–1623, 1999.

[39] H. M. Kojabadi, L. Chang, and T. Boutot, "Development of a novel wind turbine simulator for wind energy conversion systems using an inverter-controlled induction motor," *IEEE Transactions on Energy Conversion*, vol. 19, no. 3, pp. 547–552, 2004.

[40] R. Karki, P. Hu, and R. Billinton, "A simplified wind power generation model for reliability evaluation," *IEEE Transactions on Energy Conversion*, vol. 21, no. 2, pp. 533–540, 2006.

[41] R. Billinton and Y. Gao, "Multistate wind energy conversion system models for adequacy assessment of generating systems incorporating wind energy," *IEEE Transactions on Energy Conversion*, vol. 23, no. 1, pp. 163–170, 2008.

[42] L. Y. Pao and K. E. Johnson, "A tutorial on the dynamics and control of wind turbines and wind farms," in *American Control Conference*, pp. 2076–2089, IEEE, 2009.

[43] N. W. Miller, J. J. Sanchez-Gasca, W. W. Price, and R. W. Delmerico, "Dynamic modeling of ge 1.5 and 3.6 mw wind turbine-generators for stability simulations," in *IEEE Power Engineering Society General Meeting*, pp. 1977–1983, IEEE, 2003.

[44] C. Dirscherl, C. Hackl, and K. Schechner, "Modellierung und Regelung von modernen Windkraftanlagen: Eine Einführung (see https://arxiv.org/abs/1703.08661 for the English translation)," in *Elektrische Antriebe – Regelung von Antriebssystemen* (D. Schröder, ed.), ch. 24, pp. 1540–1614, Springer-Verlag, 2015.

[45] M. Schubert, "Verfahren zur Regelung einer Windenergieanlage und Windenergieanlage mit einem Rotor," 2006.

[46] A. Betz, *Wind-Energie und ihre Ausnutzung durch Windmühlen*. Vandenhoeck & Ruprecht, 1926 [Ausg. 1925].

[47] S. Heier, *Windkraftanlagen: Systemauslegung, Netzintegration und Regelung*. Vieweg+Teubner Verlag, 5. ed., 2009.

[48] C. M. Hackl, *Non-identifier based adaptive control in mechatronics: Theory and Application*. No. 466 in Lecture Notes in Control and Information Sciences, Berlin: Springer International Publishing, 2017.

[49] F. W. Koch, *Simulation und Analyse der dynamischen Wechselwirkung von Windenergieanlagen mit dem Elektroenergiesystem*. PhD thesis, Universität Duisburg-Essen, 2005.

[50] J. Slootweg, S. W. H. De Haan, H. Polinder, and W. Kling, "General model for representing variable speed wind turbines in power system dynamics simulations," *IEEE Transactions on Power Systems*, vol. 18, no. 1, pp. 144–151, 2003.

[51] J. Slootweg, H. Polinder, and W. Kling, "Initialization of wind turbine models in power system dynamics simulations," in *IEEE Porto Power Tech Proceedings*, vol. 4, (Porto. Purtugal), pp. 6 pp. vol.4–, 2001.

[52] H. Eldeeb, C. M. Hackl, and J. Kullick, "Efficient operation of anisotropic synchronous machines for wind energy systems," in *Proceedings of the 6th edition of the conference "The Science of Making Torque from Wind" (TORQUE 2016)*, Open access Journal of Physics: Conference Series 753, (Munich, Germany), 2016.

[53] F. Blaabjerg, M. Liserre, and K. Ma, "Power electronics converters for wind turbine systems," *IEEE Transactions on Industrial Applications*, vol. 48, no. 2, pp. 708–718, 2012.

[54] D. Schröder, *Leistungselektronische Schaltungen: Funktion, Auslegung und Anwendung*. Berlin: Springer-Verlag, 2012.

[55] S. Bernet, *Selbstgeführte Stromrichter am Gleichspannungszwischenkreis: Funktion, Modulation und Regelung*. Springer-Verlag, 2012.

[56] D. G. Holmes and T. A. Lipo, *Pulse Width Modulation For Power Converters*. IEEE Series on Power Engineering, Piscataway, New Jersey: IEEE Press, 2003.

[57] D. Schröder, *Leistungselektronische Schaltungen - Funktion, Auslegung und Anwendung (2. Auflage)*. Berlin: Springer-Verlag, 2008.

[58] J. Böcker, S. Beineke, and A. Bähr, "On the control bandwidth of servo drives," in *Proceedings of the 13th European Conference on Power Electronics and Applications*, (Barcelona, Spain), pp. 1–10, 2009.

[59] DIN Deutsches Institut für Normung e. V., "DIN EN 50160:2011-02: Merkmale der Spannung in öffentlichen Elektrizitätsversorgungsnetzen," 2011.

[60] C. Dirscherl, J. Fessler, C. M. Hackl, and H. Ipach, "State-feedback controller and observer design for grid-connected voltage source power converters with LCL-filter," in *Proceedings of the 2015 IEEE International Conference on Control Applications (CCA) (CCA)*, (Sydney, Australia), pp. 215–222, 2015.

[61] C. M. Hackl, "MPC with analytical solution and integral error feedback for LTI MIMO systems and its application to current control of grid-connected power converters with LCL-filter," in *Proceedings of the 2015 IEEE International Symposium on Predictive Control of Electrical Drives and Power Electronics (PRECEDE)*, (Valparaiso, Chile), pp. 61–66, 2015.

[62] D. Schröder, *Elektrische Antriebe - Regelung von Antriebssystemen (3., bearb. Auflage)*. Berlin: Springer-Verlag, 2009.

[63] C. M. Hackl, "On the equivalence of proportional-integral and proportional-resonant controllers with anti-windup," *arXiv:1610.07133*, 2016.

[64] K. J. Åström and L. Rundqwist, "Integrator windup and how to avoid it," in *Inproceedings of the American Control Conference*, (Pittsburgh, PA, USA), pp. 1693–1698, 1989.

[65] Y. Peng, D. Vranic, and R. Hanus, "Anti-windup, bumpless, and conditioned transfer techniques for PID controllers," *IEEE Control Systems Magazine*, vol. 16, no. 4, pp. 48–57, 1996.

[66] G. Kessler, "Über die Vorausberechnung optimal abgestimmter Regelkreise – Teil 3: Die optimale Einstellung des Reglers nach dem Betragsoptimum," *Regelungstechnik*, vol. 3, no. 2, pp. 40–49, 1955.

[67] C. Hackl and K. Schechner, "Non-ideal feedforward torque control of wind turbines: Impacts on annual energy production & gross earnings," *Journal of Physics: Conference Series*, vol. 753, no. 11, p. 112010, 2016.

[68] J. Mullen and J. B. Hoagg, “Wind turbine torque control for unsteady wind speeds using approximate-angular-acceleration feedback,” in *Inproceedings of the 52nd IEEE Conference on Decision and Control (CDC)*, (Florence, Italy), pp. 397–402, 2013.
[69] C. Dirscherl, C. M. Hackl, and K. Schechner, “Pole-placement based nonlinear state-feedback control of the DC-link voltage in grid-connected voltage source power converters: A preliminary study,” in *Proceedings of the 2015 IEEE International Conference on Control Applications (CCA) (CCA)*, (Sydney, Australia), pp. 207–214, 2015.
[70] K. Schechner, F. Bauer, and C. M. Hackl, “Nonlinear DC-link PI control for airborne wind energy systems during pumping mode,” in *Airborne Wind Energy: Advances in Technology Development and Research* (R. Schmehl, ed.), Springer-Verlag, 2016.
[71] V. Duindam, A. Macchelli, S. Stramigioli, and H. Bruyninckx, eds., *Modeling and control of complex physical systems: the port-Hamiltonian approach*. Berlin Heidelberg: Springer, 2009.

Table II: *Model and controller parameters for implementation and simulation.*

| Description | Symbol | Value (with unit) |
|---|---|---|
| *Turbine & gear (direct drive)* | | |
| Air density | $\varrho$ | $1.293\,\frac{\mathrm{kg}}{\mathrm{m}^3}$ |
| Turbine radius | $r_\mathrm{t}$ | $40\,\mathrm{m}$ |
| Turbine inertia | $\Theta_\mathrm{t}$ | $8.6\cdot 10^6\,\mathrm{kg\,m}^2$ |
| Power coefficient | $c_{\mathrm{p},2}(\cdot,\cdot)$ | as in (8) |
| Maximal change rate of pitch angle | $\dot{\beta}_\mathrm{max}$ | $8\,\frac{^\circ}{\mathrm{s}}$ |
| Pitch control time constant | $T_\beta$ | $0.5\,\mathrm{s}$ |
| Gear ratio | $g_\mathrm{r}$ | $1$ |
| *Permanent-magnet synchronous generator (isotropic)* | | |
| Number of pole pairs | $n_\mathrm{p}$ | $48$ |
| Stator resistance | $R_\mathrm{s}$ | $0.01\,\Omega$ |
| Stator inductance(s) | $L_\mathrm{s}^d = L_\mathrm{s}^q$ | $3.0\,\mathrm{mH}$ |
| flux linkage of permanent magnet | $\psi_\mathrm{pm}$ | $12.9\,\mathrm{V\,s}$ |
| Generator inertia | $\Theta_\mathrm{m}$ | $1.3\cdot 10^6\,\mathrm{kg\,m}^2$ |
| *Back-to-back converter* | | |
| DC-link capacitance | $C_\mathrm{dc}$ | $2.4\,\mathrm{mF}$ |
| Switching frequency | $f_\mathrm{sw}$ | $2.5\,\mathrm{kHz}$ |
| Delay | $T_\mathrm{delay} = \frac{1}{f_\mathrm{sw}}$ | $0.4\,\mathrm{ms}$ |
| *Filter & grid voltage* | | |
| Filter resistance | $R_\mathrm{f}$ | $0.1\,\Omega$ |
| Filter inductance | $L_\mathrm{f}$ | $6\,\mathrm{mH}$ |
| Grid angular frequency | $\omega_\mathrm{g} = 2\pi f_\mathrm{g}$ | $100\pi\,\frac{\mathrm{rad}}{\mathrm{s}}$ |
| Grid voltage amplitude | $\hat{u}_\mathrm{g}$ | $2.7\,\mathrm{kV}$ |
| Grid voltage initial angle | $\alpha_0$ | $0\,\mathrm{rad}$ |
| *Controller parameters* | | |
| PI current controller (37) | $k_{\mathrm{f,p}}$ | $7.5\,\Omega$ |
| (grid-side) | $k_{\mathrm{f,i}}$ | $125\,\Omega\,\mathrm{s}$ |
| | $\Delta_{\boldsymbol{\xi}_\mathrm{f}}$ | $1\cdot 10^{-3}\,\mathrm{V}$ |
| | $\widehat{u}$ | $\frac{u_\mathrm{dc}}{\sqrt{3}}$ |
| PI current controller (37) | $k_{\mathrm{s,p}}$ | $3.75\,\Omega$ |
| (machine-side) | $k_{\mathrm{s,i}}$ | $12.5\,\Omega\,\mathrm{s}$ |
| | $\Delta_{\boldsymbol{\xi}_\mathrm{s}}$ | $1\cdot 10^{-3}\,\mathrm{V}$ |
| | $\widehat{u}$ | $\frac{u_\mathrm{dc}}{\sqrt{3}}$ |
| Speed controller (38) | $k_\mathrm{p}^\star$ | $282.80\,\frac{\mathrm{kN\,m}}{\mathrm{s}^2}$ |
| | $m_{\mathrm{m,rated}}$ | $1.0419\,\mathrm{MN\,m}$ |
| DC-link voltage | $k_{\mathrm{dc,p}}$ | $-0.576\,\frac{\mathrm{A}}{\mathrm{V}}$ |
| PI controller (40) | $k_{\mathrm{dc,i}}$ | $-18.33\,\frac{\mathrm{A\,s}}{\mathrm{V}}$ |
| Phased-locked loop | $V_{\mathrm{r,pll}}$ | $20\,000\,\frac{1}{\mathrm{s}}$ |
| PI controller as in [44] | $T_{\mathrm{n,pll}}$ | $0.2\,\mathrm{ms}$ |
| Pitch angle reference | $k_{\beta,\mathrm{p}}$ | $-400.2\,\frac{^\circ\,\mathrm{s}}{\mathrm{rad}}$ |
| PI controller (39) | $k_{\beta,\mathrm{i}}$ | $-100.1\,\frac{^\circ\,\mathrm{s}^2}{\mathrm{rad}}$ |
| | $\Delta_{\xi_\beta}$ | $1\cdot 10^{-3}\,^\circ$ |
| | $\omega_{\mathrm{m,rated}}$ | $1.9195\,\frac{\mathrm{rad}}{\mathrm{s}}$ |